\documentclass[prx,showkeys,showpacs,groupedaddress,twocolumn]{revtex4-1}
\usepackage{amssymb,comment}
\usepackage{graphicx}
\usepackage{dcolumn}
\usepackage{bm}
\usepackage{amsmath}
\usepackage{subfigure}
\usepackage{float}
\usepackage{color}
\usepackage{txfonts}
\usepackage[colorlinks,linkcolor=blue,anchorcolor=blue,citecolor=blue,bookmarksnumbered]{hyperref}

\DeclareMathOperator{\arccot}{arccot}

\newcommand{\prt}{\partial}
\newcommand{\cG}{\mathcal{G}}
\newcommand{\cR}{\mathcal{R}}
\newcommand{\tk}{\widetilde{k}}
\newcommand{\om}{\omega}
\newcommand{\al}{\alpha}
\newcommand{\tom}{\widetilde{\omega}}
\newcommand{\tal}{\widetilde{\alpha}}

\begin{document}
\title{Accessing and Manipulating Dispersive Shock Waves in a Nonlinear and Nonlocal Rydberg Medium}
\author{Chao Hang$^{1,4,6}$}
\author{Zhengyang Bai$^{1}$}
\author{Weibin Li$^{2,5}$}
\thanks{Corresponding author. weibin.li@nottingham.ac.uk}
\author{Anatoly M. Kamchatnov$^{3}$}
\author{Guoxiang Huang$^{1,4,6}$}
\thanks{Corresponding author. gxhuang@phy.ecdnu.edu.cn}
\affiliation{$^{1}$State Key Laboratory of Precision Spectroscopy, East China Normal University, Shanghai 200062, China}
\affiliation{$^{2}$School of Physics and Astronomy, University of Nottingham, Nottingham NG7 2RD, England, United Kingdom}
\affiliation{$^{3}$Institute of Spectroscopy, Russian Academy of Sciences, Troitsk, Moscow 108840, Russia}
\affiliation{$^{4}$NYU-ECNU Institute of Physics, New York University at Shanghai, Shanghai 200062, China}
\affiliation{$^{5}$Centre for the Mathematics and Theoretical Physics of Quantum Non-Equilibrium Systems, University of Nottingham, Nottingham NG7 2RD, England, United Kingdom}
\affiliation{$^{6}$Collaborative Innovation Center of Extreme Optics, Shanxi University, Taiyuan, Shanxi 030006, China}

\date{\today}

\begin{abstract}
Dispersive shock waves (DSWs) are fascinating wave phenomena occurring in media when nonlinearity overwhelms dispersion (or diffraction). Creating DSWs with low generation power and realizing their active controls is desirable but remains a longstanding challenge. Here, we propose a scheme to generate weak-light DSWs and realize their manipulations in an atomic gas involving strongly interacting Rydberg states under the condition of electromagnetically induced transparency (EIT).  We show that for a two-dimensional (2D) Rydberg gas a weak nonlocality of optical Kerr nonlinearity can significantly change the edge speed of DSWs, and induces a singular behavior of the edge speed and hence an instability of the DSWs. However, by increasing the degree of the Kerr nonlocality, the singular behavior of the edge speed and the instability of the DSWs can be suppressed.
We also show that in a 3D Rydberg gas, DSWs can be created and propagate stably when the system works in the intermediate nonlocality regime. Due to the EIT effect and the giant nonlocal Kerr nonlinearity contributed by the Rydberg-Rydberg interaction, DSWs found here have extremely low generation power. In addition, an active control of DSWs can be realized; in particular, they can be stored and retrieved with high efficiency and fidelity through switching off and on a control laser field. The results reported here are useful not only for unveiling intriguing physics of DSWs but also for finding promising applications of nonlinear and nonlocal Rydberg media.


\end{abstract}

\date{\today}

\maketitle

\section{Introduction}

A shock wave is a typical propagating disturbance characterized by an abrupt, nearly discontinuous change in the characteristics of material medium.
Dispersive shock waves (DSWs) are widespread phenomena occurring in various physical systems, including fluids~\cite{Peregrine1966,Smyth1988,Maiden2016},
plasmas~\cite{Taylor1970,Ikezi1973,Romagnani2008}, Bose-Einstein condensates (BECs)~\cite{Damski2004,Kamchatnov2004,Garcia2004,Simula2005,Hoefer2006,Chang2008}, electron gases~\cite{Bettelheim2006}, and optical media~\cite{Rothenberg1989,El2007,Hoefer2007,Ghofraniha2007,Wen2007,Conti2009,
Ghofraniha2012,Garnier2013,Gentilini2013,Gentilini2014,Gentilini2015PRA,
Gentilini2015SR,Xu2015,Braidotti2016,Wetzel2016,Xu2016OL,Xu2016Phy,
Xu2017,Marcucci2019NC,Marcucci2020,Bienaime2021}.
In optics, if propagation distance is short enough and a laser pulse (or beam) in a nonlinear medium can reasonably be described by disregarding effects of dissipation and dispersion (or diffraction), an initially smooth laser pulse (or beam) will steepen rapidly during propagation because of the nonlinear effect and it will arrive at a point of gradient catastrophe, known as wave breaking. If the dissipation is negligibly small, after the occurrence of the wave breaking the laser pulse (or beam) will acquire an oscillatory structure by the interplay between the nonlinearity and dispersion (or diffraction), called a dispersive shock wave (DSW). 
Otherwise, if the dissipation dominates over the dispersion/diffraction, the laser pulse (or beam) will acquire only a smooth front without any
oscillations, alias as dissipative shock wave.

On the other hand, optical materials with \textit{nonlinearities} and \textit{nonlocalities} are of great interests due to their intriguing physics and practical applications. 
Particularly, Rydberg atomic gases~\cite{Gallagher2008} working under the condition of electromagnetically induced transparency (EIT)~\cite{Fleischhauer2005} possess
many unique properties, which include: (i)~the optical absorption due to the resonance between optical fields and atoms can be greatly suppressed via the EIT~\cite{Mohapatra2007}, a quantum de-construction interference effect induced by a control laser field; (ii)~they can map the strong and long-range interaction (i.e. the Rydberg-Rydberg interaction) between atoms in Rydberg states into the strong and long-range interaction between photons~\cite{Peyronel2012}, resulting in a giant and nonlocal optical Kerr nonlinearity; (iii)~they are configurable and controllable in an active way due to the existence of many tunable parameters~\cite{Maxwell2013a}, such as atomic levels, detuning, and laser intensities, etc. Based on these striking features, Rydberg-EIT systems have become an excellent platform for the research of quantum and nonlinear optics in strongly interacting atomic ensembles~\cite{Sevincli2011,Pritchard2013,Gorshkov2013,Firstenberg2016,Murray2016,Bai2019,Bai2020}, and have promising applications in many fields such as high precision measurement and quantum information processing~\cite{Saffman2010,Murray2017,Ding2022}.

In many situations light propagation in media with a local Kerr nonlinearity can be described by a nonlinear envelope equation, i.e. the local nonlinear Schr\"{o}dinger equation (NLSE), which is completely integrable and can be solved exactly by the inverse scattering transform~\cite{Ablowitz1991}. A general approach for describing DSWs in local nonlinear media was developed by Gurevich and Pitaevskii~\cite{gp-73}, which is based on the Whitham modulation theory of nonlinear waves~\cite{whitham-65,whitham-74}. In this approach, DSWs are approximated by modulated periodic-wave solutions, and the evolution of solution variables is governed by the Whitham modulation equations (see, e.g., \cite{fl-86,pavlov-87,gk-87,eggk-95} and review papers \cite{eh-16,kamch-21c}).
However, for systems with nonlocal Kerr nonlinearities, the nonlinear envelope equation will be modified into a nonlocal NLSE (NNLSE), which
leads to significant consequences. Especially, the NNLSE is non-integrable and can not be solved by using the inverse scattering transform method. Yet, the Gurevich-Pitaevskii method is still applicable and the main characteristics of DSWs can still be attained from the restricted Whitham equations~\cite{gm-84,el-05,egs-06,kamch-19,kamch-20}.

In this work, we propose a scheme to generate DSWs at weak-light level and realize their active manipulations. The system under study consists of a cold Rydberg atomic gas with a ladder-type energy-level configuration under the condition of EIT (i.e. Rydberg-EIT), which possesses a giant nonlocal Kerr nonlinearity with vanishing absorption. We derive a NNLSE governing the propagation of a probe laser beam, and investigate the formation, propagation, and control of various types of DSWs in two- and three-dimensional Rydberg gases.

Firstly, we show that DSWs can be created in a two-dimensional (2D) Rydberg gas~\cite{note-definiton-2D}. We find that even the existence of a very weak nonlocality of the Kerr nonlinearity (i.e. in the case that the Rydberg blockade radius~\cite{note-blockade} is much smaller than the probe beam radius) can significantly change the edge speeds of DSWs. The weak nonlocality can also make the edge speeds display a singular behavior when the local sound speed $c_s$ of the light fluid is in the vicinity of the critical value $c_{cr}$. Furthermore, it can induce an instability of DSWs, which emerges from the small-amplitude edge but not the soliton-amplitude edge, when $c_s\ge c_{cr}$. However, for a moderate degree of nonlocality (i.e. in the case that the Rydberg blockade radius is of the same order with that of the probe beam radius), the increase of the edge speeds becomes much slower than that in the weak nonlocality regime when $c_s$ increases. In this situation, the singular behavior of the edge speeds vanishes and the instability of DSWs is thoroughly suppressed.

Secondly, we show that isotropic and anisotropic DSWs can be created in a 3D Rydberg gas, where the wave breaking occurs in both two transversal spatial dimensions.
Such DSWs are stable during propagation if the system works in the regime of intermediate nonlocality. Moreover, spatiotemporal DSWs can also be excited in the intermediate nonlocality regime, for which the wave breaking occurs in two transversal spatial dimensions and one time dimension. We demonstrate that all the DSWs found here have extremely low generation power ($\le 10$ nano-watts). In addition, such DSWs can be manipulated actively via adjusting system parameters; in particular, they can be easily stored and retrieved with high efficiency and fidelity through switching off and on a control laser field.

We would like to emphasize that the interesting properties of DSWs are stemmed from the EIT effect and the giant nonlocal Kerr nonlinearity contributed by the Rydberg-Rydberg interaction. The results reported in this work differ considerably from the nonlocality effects in other media as, for example, nematic liquid crystals (see \cite{es-16} and references therein) and they are useful not only for unveiling novel physics of nonlinear and nonlocal media but also for promising applications for optical information processing and transformation.

The remainder of the article is arranged as follows. In Sec.~\ref{section2}, we describe the physical model and present the derivation of the NNLSE
governing the nonlinear propagation of the probe laser beam. In Sec.~\ref{section3}, we study the sound propagation in weak nonlocality regime. In Sec.~\ref{section4} and Sec.~\ref{section5}, we present detailed results on the formation and propagation of DSWs in multi-dimensional gases. 
In Sec.~\ref{section6}, we consider the storage and retrieval of DSWs. Finally, Sec.~\ref{section7} contains summary on the main results obtained in this work.

\section{Model and envelope equation}\label{section2}

\subsection{Physical model}

\begin{figure}
\centering
\includegraphics[width=\columnwidth]{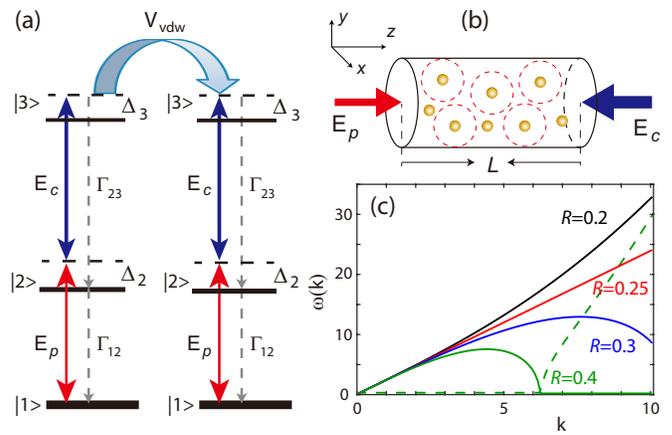}
\caption{{\footnotesize (a)~Energy-level diagram and excitation scheme of the Rydberg gas. $|\alpha\rangle$ ($=1,\,2,\,3$) are atomic quantum states; $\Delta_{2}$ and $\Delta_{3}$ are respectively one- and two-photon detuning;
$\Gamma_{12}$ and $\Gamma_{23}$ are respectively decay rates of $|2\rangle$ and $|3\rangle$; $E_p$ and $E_c$ are respectively the probe and control fields.
$V_{\rm vdw}({\bf r'}-{\bf r})=-\hbar C_6/|{\bf r'}-{\bf r}|^6$ is the van der Waals potential between the Rydberg atoms at positions ${\bf r}$ and ${\bf r'}$, respectively. (b)~Geometry for possible experimental arrangement, where dots denote ground atoms, dashed circles denote Rydberg blockade radius, and $L$ is the sample length. (c)~Linear dispersion relation $\omega=\omega(k)$ [given by Eq.~(\ref{dispersion})] as a function of wavenumber $k$ for $\rho_b=1$ and ${\cal G}=5.7$, with ${\cal R}=0.2$ (black solid line), ${\cal R}=0.25$ (red solid line; it is a straight line), ${\cal R}=0.3$ (blue solid line), and ${\cal R}=0.4$ (green solid line). See text for details.
}}
\label{fig1}
\end{figure}

We start to consider a cold atomic gas with a ladder-type three-level configuration [see Fig.~\ref{fig1}(a)], where $|1\rangle$, $|2\rangle$, and $|3\rangle$ denote respectively the ground, intermediate, and high-lying Rydberg state.  A weak probe laser field $\mathbf E_p$ (with angular frequency $\omega _{p}$ and wavenumber $k_p=\omega_p/c$) couples the transition $|1\rangle\leftrightarrow |2\rangle$, and a strong control laser field $\mathbf E_c$ (with angular frequency $\omega _{c}$ and wavenumber $k_c=\omega_c/c$) couples the transition $|2\rangle\leftrightarrow |3\rangle$.
%
%
The total electric-field vector reads
$\mathbf{E}=\mathbf{E}_p+\mathbf{E}_c
=\sum_{l=p,c}{\mathbf{e}_{l}\mathcal{E}}_{l}\exp
[i({\bf k}_l\cdot {\bf r}-\omega_l t)]+{\rm c.c}.$, where c.c. represents the complex conjugate of the preceding term; $\mathbf{e}_{p}$ and $\mathcal{E}_{p}$
($\mathbf{e}_{c}$ and $\mathcal{E}_{c}$) are respectively the polarization unit vector and envelope of the probe (control) field. For avoiding residue Doppler effect, the probe and control fields are assumed to counter-propagate along the $z$-axis [Fig.~\ref{fig1}(b)], and hence we have ${\bf k}_{p}=k_{p}{\bf e}_z$ and ${\bf k}_c=-k_c{\bf e}_z$, with ${\bf e}_z$ the unit vector along the $z$ direction.

The Hamiltonian of the system under electric-dipole and rotating-wave approximations is given by $\hat{H}={\cal N}_a \int d^3 {\bf r}\, \hat{\cal H}$, where ${\cal N}_a$ is the atomic density and $\hat{\cal H}$ is the
Hamiltonian density with the form
\begin{align}\label{Hamiltonian}
{\cal \hat{H}}=-&\hbar\sum_{\alpha
=2}^{3}{\Delta_{\alpha}\hat{S}_{\alpha\alpha}\left({\bf r},t \right)}- \hbar\left[\Omega_{p}\hat{S}_{12}+\Omega_c\hat{S}_{23} +{\rm h.c.} \right] \notag \\
&+\mathcal N_{a} \int{d^3 {\bf r}^{\prime}\hat{S}_{33}({\bf r}^{\prime},t) V_{\rm vdw} ({\bf r}^{\prime}-{\bf r}) \hat{S}_{33} ({\bf r},t )}.
\end{align}
Here, $\Delta_2=\omega_p-(E_2-E_1)/\hbar$ and $\Delta_3=\omega_p+\omega_c-(E_3-E_1)/\hbar$ are respectively one- and two-photon detuning, with $E_{\alpha}$ the eigenenergy of the atomic state $|\alpha\rangle$;
$\hat{S}_{\alpha\beta}=|\beta\rangle \langle \alpha| \exp\{i[({\bf k}_\beta-{\bf k}_\alpha)\cdot {\bf r}-(\omega_\beta-\omega_\alpha+\Delta_\beta-\Delta_\alpha)t ]\}$ ($\alpha,\,\beta=$1, 2, 3) are atomic transition operators associated with the states $\alpha$ and $\beta$, satisfying the commutation relation $\left[\hat{S}_{\alpha \beta}(\mathbf{r},t),\hat{S}_{\mu\nu}(\mathbf{r}^\prime,t)\right]
=(1/{\cal N}_a)\delta({\bf r}-{\bf r}^{\prime})\left[\delta_{\alpha\nu}\hat{S}_{\mu \beta}(\mathbf{r}^\prime,t)-\delta_{\mu \beta}\hat{S}_{\alpha\nu}(\mathbf{r}^\prime,t)\right]$;
$\Omega_p={(\mathbf
e_p\cdot\mathbf p_{12}) \mathcal{E}_p}/(2\hbar)$ and $\Omega_c={(\mathbf e_c\cdot\mathbf p_{23}) \mathcal{E}_c}/(2\hbar)$ are respectively half Rabi frequencies of the probe and control fields, with $\mathbf
p_{\alpha\beta}$ the electric-dipole matrix elements associated with the transition $|\beta\rangle\leftrightarrow |\alpha\rangle$.
The last term on the right-hand side of Eq.~(\ref{Hamiltonian}) is stemmed from the strongly interacting Rydberg states, where $V_{\rm vdw}({\bf r'}-{\bf r})=-\hbar C_6/|{\bf r'}-{\bf r}|^6$ is the van der Waals (vdW) interaction potential between the Rydberg atoms located at the positions
${\bf r'}$ and ${\bf r}$~\cite{Gallagher2008}, with $C_6$ the dispersion parameter determined by the characteristics of atoms.

Under slowly varying envelope approximation, the Maxwell equation governing the propagation of the probe field is reduced into
\begin{equation}\label{Max}
i\left( \frac{\partial}{\partial z} + \frac{1}{c} \frac{\partial}{\partial t}  \right) \Omega_p+\frac{1}{2k_p}\nabla_{\perp}^{2}\Omega_p+ \frac{k_p}{2}\chi_{p}\Omega_p=0,
\end{equation}
where $\nabla_{\perp}^{2}=\partial_{xx}+\partial_{yy}$ is transverse Laplacian operator and $\chi_{p}=\mathcal{N}_{a}(\mathbf{e}_{p}\cdot \mathbf{p}_{12})^{2}\rho _{21}/(\varepsilon_{0}\hbar \Omega_{p})$ is
probe-field susceptibility.
The dynamics of the atomic gas is controlled by the optical Bloch equation
\begin{equation}\label{Bloch0}
\frac{\partial {\hat \rho}}{\partial t}=-\frac{i}{\hbar}\left[\hat{H}, {\hat \rho}\right]- \Gamma\,[{\hat\rho}],
\end{equation}
where ${\hat \rho}$ is a $6\times 6$ density matrix, with the matrix element $\rho_{\alpha\beta}\equiv \langle {\hat S}_{\alpha\beta}\rangle$~\cite{note-average};
$\Gamma$ is a $6\times 6$ relaxation matrix describing the spontaneous emission and dephasing of the atoms. The explicit expression of Eq.~(\ref{Bloch0}) is presented in Appendix~\ref{appendixA}.

\subsection{Nonlinear envelope equation}

Since in our consideration the probe field is much weaker than the control field, a perturbation method can be adopted to solve the Maxwell-Bloch (MB) equations (\ref{Max}) and (\ref{Bloch0}). When solving the MB equations,
a key point is how to give an appropriate theoretical approach on the many-body correlations. 
From Eq.~(\ref{Bloch}) in the Appendix~\ref{appendixA}, one sees that
the equations for the one-body correlations $\rho_{\alpha\beta}({\bf r},t)$ involve two-body correlations $\rho_{33,3\alpha}({\bf r}',{\bf r},t)=\langle\hat{S}_{33}({\bf r}',t)\hat{S}_{3\alpha}({\bf r},t)\rangle$ ($\alpha=1,2$), so one must solve the equations for the two-body correlations, which, however, involve three-body correlations, and so on. Such an infinite equation chain (i.e. the BBGKY hierarchy) can be solved by the reductive density matrix expansion developed in Refs~\cite{Bai2019,Bai2016}. Based on this approach,
we obtain the following (3+1)D~\cite{note-definiton-(3+1)D} NNLSE
\begin{align}\label{NLS}
&i\left(\frac{\partial}{\partial z}-\frac{1}{V_g}\frac{\partial}{\partial t}\right)\Omega_p - \frac{K_2}{2}\frac{\partial^2 \Omega_p}{\partial t^2} + \frac{1}{2k_{p}}\nabla _ \bot ^2\Omega_p + \notag\\
&\quad \int d^3{\bf r^{\prime}} G({\bf r^{\prime}- r}){\left| {{\Omega_p({\bf r^{\prime}},t)}} \right|}^2  {\Omega_p({\bf r},t)}= -iA\Omega_p ,
\end{align}
where
\begin{equation}\label{Vg}
V_g=\left(\frac{1}{c}+\kappa_{12}\frac{|\Omega_c|^2+d_{31}^2}{D_1^2}\right)^{-1},
\end{equation}
characterizes the group velocity of the probe-field envelope, with $c$ the light speed in vacuum ($c\approx2.99\times10^{10}$ cm s$^{-1}$) and $D_1=|\Omega_c|^2-d_{21}d_{31}$, with $d_{\alpha 1}=\Delta_\alpha+i\Gamma_{\alpha}/2$ ($\Gamma_{\alpha}$ is the decay rate of the state $|\alpha\rangle$; $\alpha=2$, 3). The second term on the left-hand side describes the group-velocity (or second-order) dispersion of the system, with the coefficient $K_2$ given by
\begin{equation}\label{K2}
K_2=2\kappa_{12}\frac{(d_{21}+2d_{31})|\Omega_c|^2+d_{31}^3}{D_1^3}.
\end{equation}

The last term on the left-hand side is contributed by the Kerr nonlinearity originated from the Rydberg-Rydberg interaction, in which the nonlocal response function $G({\bf r^{\prime}- r})$ (describing the collective photon-photon interaction) owns the form
\begin{eqnarray}\label{G}
&& G({\bf r^{\prime}- r}) = -\frac{k_p(\mathbf{e}_{p}\cdot \mathbf{p}_{12})^{2}}{2\epsilon_{0}\hbar}\frac{\mathcal{N}_{a}^2d_{21}}{D_1}\,
R({\bf r}'-{\bf r})
V_{\rm vdw}({\bf r}'-{\bf r}),
\end{eqnarray}
with
\begin{eqnarray}\label{R}
&& R({\bf r}'-{\bf r})=\frac{\sum\limits_{m=0}^{2}P_{m}V_{\rm vdw}({\bf r}'-{\bf r})^m}{\sum\limits_{n=0}^{3}Q_{n}V_{\rm vdw}({\bf r}'-{\bf r})^n}.
\end{eqnarray}
Here, $P_n$ and $Q_m$ are functions of system parameters (including $\Delta_{\alpha}$, $\gamma_{\alpha\beta}$, and $\Omega_c$), whose explicit expressions are very cumbersome and hence omitted here. Because the dephasing in the system is much smaller than the spontaneous emission, Eq.~(\ref{R}) can be further simplified to be
\begin{eqnarray}
&& R({\bf r}'-{\bf r})
\approx-\frac{2(d_{21}+d_{31})|\Omega_c|^2\Omega_c/|D_1|^2}{2d_{21}|\Omega_c|^2
+D_2[2d_{31}-V_{\rm vdw}({\mathbf{r}'-\mathbf{r}})]},
\end{eqnarray}
with $D_2=|\Omega_c|^2-d_{21}(d_{21}+d_{31})$. It is due to the contribution of $R({\bf r}'-{\bf r})$ that makes the response function $G({\bf r})$ have a soft-core profile near $|{\bf r}|=0$~\cite{Sevincli2011}.

The term on the right-hand side of (\ref{NLS}) describes the linear optical absorption of the medium, where
$A=2\kappa_{12}{\rm Im}\left(d_{31}/D_1\right)$. Since under the EIT and large one-photon detuning conditions, i.e. $|\Omega_c|^2\gg \gamma_{21}\gamma_{31}$ and $|\Delta_{2}|\gg\gamma_{21}$, $A$ is very small in comparison with other coefficients~\cite{Fleischhauer2005}, the term on the right hand side of Eq.~(\ref{NLS}) will be neglected in the following discussions except in Sec.~\ref{section6}.

The probe-field susceptibility can be expanded as the form of $\chi _{p}\approx
\chi_p ^{(1)}+\int d^{3} \mathbf{r}^{\prime} \chi_p ^{(3)}\left(\mathbf{r}^{\prime}-\mathbf{r}\right)\left|
\mathcal{E}_{p}\left(\mathbf{r}^{\prime}\right)\right|^{2}$, where $\chi_p ^{(1)}$ and $\chi_p ^{(3)}$ are respectively the linear and third-order nonlinear optical susceptibilities. The relation between $\chi_p ^{(3)}$ and the nonlinear response function $G({\bf r^{\prime}- r})$ is given by
\begin{align}
\chi ^{(3)}({\bf r^{\prime}- r})=\frac{2(\mathbf{e}_{p}\cdot \mathbf{p}_{12})^{2}}{k_{p}\hbar ^{2}}G({\bf r^{\prime}- r}).
\end{align}
Two key features of the nonlinear susceptibility $\chi^{(3)}$ are the following: (i)~it is position-dependent, i.e. nonlocal in space; (ii)~it can be enhanced greatly; in fact, it can reach to the order of magnitude of $10^{-9}~\mathrm{m^{2}V^{-2}}$, which is more than 11 orders larger than that obtained by using common nonlinear optical materials, such as optical fibers~\cite{Bai2016,Mu2021}. These unique features are rooted from the strong and long-ranged Rydberg-Rydberg interaction.

\section{Sound propagation in weak nonlocality regime}\label{section3}

We first investigate the sound propagation based on the hydrodynamical representation of the NNLSE (\ref{NLS}). In order to extract analytical results, we consider the case of DSWs under some realistic physical conditions.

\subsection{Characteristic quantities and systemic parameters}\label{section3A}

If the probe beam is rather extended in the $y$ direction, one can reduce Eq.~(\ref{NLS}) into the dimensionless form
\begin{align}\label{NLS1}
&i\left(\frac {\partial }{\partial \zeta}+\frac{1}{\lambda}\frac{\partial }{\partial \tau}\right)U+{\cal D}\frac{\partial^2 U}{\partial \tau^2}+\frac{1}{2}\frac{\partial^2 U}{\partial \xi^2}  \notag\\
&\quad + {\cal G}\int d\xi' g(\xi'-\xi)|U(\xi',\zeta,\tau)|^2U=0,
\end{align}
with the new variables defined by $U=\Omega_p/\Omega_0$, $\xi=x/R_0$, $\zeta=z/L_{\rm diff}$, $\tau=t/\tau_0$,
$\lambda=V_g\tau_0/L_{\rm diff}$,
${\cal D}=L_{\rm diff}/L_{\rm disp}$, and
${\cal G}=L_{\rm diff}/L_{\rm nlin}$.
Here, $\Omega_0$ is maximum half Rabi frequency of the probe field, $R_0$ is beam radius, $\tau_0$ is pulse duration, $L_{\rm diff}=\omega_{p}R_0^2/c$ is characteristic diffraction length, $L_{\rm disp}=-2\tau_0^2/K_2$ is characteristic dispersion length, and $L_{\rm nlin}=1/(G_0|\Omega_0|^2)$ is characteristic nonlinearity length, the constant $G_0=\int d^3{\bf r} |G({\bf r})|$.
The dimensionless nonlocal response function is defined by $g(\xi'-\xi)=(R_0/G_0)\int dy dz G(\xi'-\xi,y,z)$, obeying the normalization condition $\int d\xi\, |g(\xi)|=1$. Eq.~(\ref{NLS1}) is a (2+1)D NNLSE with $\zeta$, $\tau$, and $\xi$ as independent variables.

We assume that the probe field is sought with the form
\begin{equation}
U(\zeta,\tau,\xi)=p(\zeta,\tau)\,u(\zeta,\xi),
\end{equation}
where  $p(\zeta,\tau)$ is a Gaussian wavepacket propagating along the $z$ direction with the group velocity $V_g$, i.e.
\begin{equation}\label{wave_packet}
p(\zeta,\tau)=\frac{1}{\sqrt[4]{2\pi\rho_0^2}}e^{-(\zeta-\lambda\tau)^2/(4\rho_0^2)}
=\frac{1}{\sqrt[4]{2\pi\rho_0^2}}e^{-(z-V_gt)^2/(4\rho_0^2L_{\rm diff}^2)},
\end{equation}
with $\rho_0$ a free real parameter. Since this wavepacket is a solution of the equation $i\left[\partial/\partial\zeta +(1/\lambda)\partial/\partial\tau\right]p=0$, Eq.~(\ref{NLS1}), after integrating over the variable $\tau$, becomes
\begin{equation}\label{NLS12}
i\frac{\partial u}{\partial \zeta}+\frac{1}{2}\frac{\partial^2 u}{\partial \xi^2} + {\cal G}\int d\xi' g(\xi'-\xi)\,|u(\xi',\zeta)|^2\,u=0,
\end{equation}
which governs the propagation of $u$ if the group-velocity dispersion can be neglected (i.e. the dimensionless dispersion parameter ${\cal D}\ll 1$). It is a (1+1)D NNLSE with $\zeta$ and $\xi$ as independent variables~\cite{note-2D-NNLSE}.

Since the above analysis applied to the Rydberg-EIT system is rather general, we will consider laser-cooled $^{87}$Rb atomic gas as an example. The atomic levels are selected to be $|1\rangle=|5 S_{1/2}\rangle$,
$|2\rangle=|5 P_{3/2}\rangle$, and $|3\rangle=|n S_{1/2}\rangle$, with the spontaneous decay rates $\Gamma_{2}\approx2\pi\times6$ MHz and $\Gamma_{3}\approx2\pi$ kHz. The value of the dispersion parameter $C_6$ depends on the principal quantum number $n$; when $n=30$, $C_6\approx-2\pi\times68$ MHz $\mu$m$^6$. The half Rabi frequency of the control field is chosen as $\Omega_c=10$ MHz while the density of the atomic gas is chosen as $\mathcal{N}_a=1.0\times10^{10}$ cm$^{-3}$. The one- and two-photon detuning are respectively taken to be $\Delta_2=60$ MHz and $\Delta_3=-0.2$ MHz, by which the system approximately works under the condition of EIT and only a small part of atoms are excited into the Rydberg state, avoiding a significant probe-field absorption due to the Rydberg blockade effect.

By using the above parameters, we obtain the numerical value of the group velocity of the probe pulse, i.e.  $V_g\approx1.5\times10^{-5}\,c$. Such an ultraslow propagation velocity of the probe-field pulse comes from the EIT effect induced by the control field. If choosing the probe beam radius $R_0\approx 7.7$ $\mu$m, time duration $\tau_0\approx1.6$ $\mu$s, and
maximum half Rabi frequency of the probe field $\Omega_0\approx5$ MHz, we obtain $L_{\rm diff}\approx0.47$ mm, $L_{\rm nlin}\approx0.11$ mm, and $L_{\rm disp}\approx2.1$\,cm
~(see Table~\ref{Table1}).
\begin{table*}
\setlength{\tabcolsep}{3mm}
\caption{Important parameters and characteristic lengths for the generation of DSWs in $^{87}$Rb atoms. For more details, see the text.}
{\begin{tabular}{ccccccccc} \hline
& & & \\[-6pt]
Parameters & $R_0$ & $R_b$ & $\tau_0$ & $\Omega_0$ & $V_g$ & $L_{\rm diff}$ & $L_{\rm disp}$ & $L_{\rm nlin}$ \\ \hline
& & & \\[-6pt]
Values & 7.7 ($\mu$m) & 2.6 ($\mu$m) & 1.6 ($\mu$s) & 5.0 (MHz) & $1.5\times10^{-5}\,c$ & 0.47 (mm) & 2.1 (cm) & 0.11 (mm) \\ \hline\hline
& & & \\[-6pt]
Parameters & ${\cal D}$ & ${\cal G}$ & $\sigma$ & ${\cal R}$ & $c_{cr}$ & & & \\ \hline
Values & 0.02 & 4.1 & 0.3 & 0.1 & 3.16 & & & \\ \hline
& & & \\[-6pt]
\end{tabular}} \label{Table1}
\end{table*}
Thereby, the dimensionless coefficients of Eq.~(\ref{NLS1}) are given by ${\cal G}\approx4.1$ and ${\cal D}\approx 0.02\ll 1$, which means that the second-order dispersion is indeed negligible.

With the given parameters, the dimensionless nonlocal response function $g(\xi'-\xi)$ given by Eq.~(\ref{NLS12}) can be written as
\begin{equation}\label{response}
g(\Delta\xi) \approx  -g_0\int dydz \left\{g_1+\frac{g_2}{\sigma^6}\left[\Delta\xi^2+\frac{(y^2+z^2)}{R_0^2}\right]^3\right\}^{-1},
\end{equation}
where $\Delta\xi=\xi'-\xi$, $g_0\approx1.26$, $g_1=1+i\,0.28$, and $g_2=1.5\times10^{-3}$. In Eq.~(\ref{response}), $\sigma$ is defined as
\begin{equation}
\sigma=R_b/R_0,
\end{equation}
which characterizes the {\it nonlocality degree} of the Kerr nonlinearity. Here $R_b=(|C_6/\delta_{\rm EIT}|)^{1/6}$ is the Rydberg blockade radius, with $\delta_{\rm EIT}=|\Omega_c|^2/|\Delta_2|$ denoting the linewidth of EIT transmission spectrum for $|\Delta_2|\gg\Gamma_{2}$. Using the parameters given above, we have $R_b\approx2.6\,\mu$m. Up to this point, we can draw the following conclusions: (i)~under the condition that
$\Delta_2$ is much larger than $\Gamma_{2}$ and $\Delta_3$ (i.e. $\Delta_2 \gg \Gamma_2,\,\Delta_3$), the imaginary part of the nonlocal response function $g(\xi'-\xi)$ is much smaller than the corresponding real part; (ii)~The interaction between photons is repulsive, thus the Kerr nonlinearity obtained is of the type of {\it defocusing}, i.e. $g(\xi'-\xi)<0$, which is crucial for the formation of DSWs.

\subsection{The envelope equation in weak nonlocality regime}

\begin{figure}
\centering
\includegraphics[width=0.85\columnwidth]{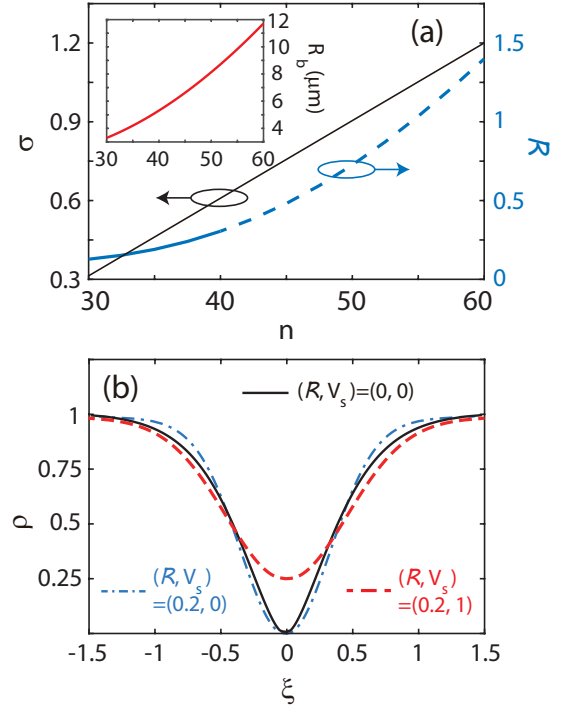}
\caption{{\footnotesize (a)~The nonlocality degree $\sigma$ (thin black line) and the intensity diffraction parameter ${\cal R}$ (thick blue line) as functions of the principal quantum number $n$ of the atoms. The dashed part in the curve of ${\cal R}$ means that ${\cal R}$ is not applicable because the system is outside of the weak nonlocality regime (${\cal R}\gtrsim0.3$). Inset: The blockade radius $R_b$ as a function of the principal quantum number $n$.
(b)~Intensity profile $\rho$ of the dark soliton  as  a function of $\xi$ for $({\cal R},V_s)=(0,0)$
(black solid line), (0.2,0) (cyan dash-dotted line), and (0.2,1) (red dashed line), respectively.}}
\label{fig2}
\end{figure}

If the nonlocality degree of the Kerr nonlinearity is weak ($\sigma\ll1$), the width of the nonlocal response function $g(\xi'-\xi)$ is finite but much narrower than the width of the probe intensity $|u(\xi',\zeta)|^2$. Hence we can expand $|u(\xi',\zeta)|^2$ around $\xi'=\xi$ in the integral of Eq.~(\ref{NLS12}), leading to the equation
\begin{equation}\label{NLS2}
i \frac {\partial u}{\partial \zeta}+\frac{1}{2}\frac{\partial^2 u}{\partial \xi^2} - {\cal G}|u|^2\,u - {\cal R}\frac {\partial^2 |u|^2}{\partial \xi^2}u=0,
\end{equation}
with ${\cal R}=-({\cal G}/2)\int \xi^2 g(\xi) d\xi$\, (since $g(\xi'-\xi)<0$, ${\cal R}>0$), referred to as the intensity diffraction parameter.
The above equation also implies that a weakly nonlocal Kerr nonlinearity (characterizing by the last term on the left-hand side of the equation) plays a role of an intensity diffraction for the probe beam, which may bring new and interesting phenomena to DSWs~\cite{notediffraction}.
We want to emphasize that parameters $\sigma$, ${\cal R}$, and $R_b$ depend on Rydberg states through principal quantum number $n$.  Shown in Fig.~\ref{fig2}(a) are
%
%
$\sigma$, ${\cal R}$, and $R_b$ as functions of $n$, given by the solid black curve, the solid-dashed blue curve, and the dashed red curve in the inset, respectively.
From the figure we see that, when $n\lesssim 40$, ${\cal R}$ takes a small value (say, ${\cal R}\lesssim 0.2$). Note that the situation $n> 40$ is outside of  the weak nonlocality regime, given by the dashed-line segment in the curve of ${\cal R}$. Importantly, these data provide an efficient tool to tune the dynamics of the optical field in different regimes.

By writing Eqs.~(\ref{NLS12}) and (\ref{NLS2}) into the momentum space, 
a relation between the intensity diffraction parameter ${\cal R}$ and the nonlocality degree $\sigma$ can be established in the weak nonlocality regime, given by
\begin{equation}\label{R-S}
{\cal R}\approx{\cal G}\sigma^2/4,
\end{equation}
which can be further reduced to ${\cal R}\approx\sigma^2$ with the system parameters of $^{87}$Rb atomic gas.


We also remark that if the Rydberg atomic medium works in {\em strong} nonlocality regime ($\sigma\gg 1$), where the width of the nonlocal response function $g(\xi'-\xi)$ is much wider than the width of the probe intensity
$|u(\xi',\zeta)|^2$, one can expand $g(\xi'-\xi)$ around $\xi'=\xi$ in the integral of Eq.~(\ref{NLS12}) and hence the equation can be simplified as
$i\partial u/\partial \zeta+(1/2)\partial^2 u/\partial \xi^2 + {\cal G}(S_0+S_1\xi+S_2\xi^2)u=0$.
Here, the parameters $S_m=d^m g(\xi)/d \xi^m |_{\xi=0}\,P/m!$ ($m=0,\,1,\,2$), with $P=\int |u(\xi,\zeta)|^2 d\xi$ the light power of the probe field. This equation can be solved by using a recently developed technique based on the time asymmetric quantum mechanics~\cite{Gentilini2015PRA,Gentilini2015SR,Braidotti2016,Marcucci2019APX,
Marcucci2020}. However, such technique is not applicable in the weak ($\sigma\ll 1$) and intermediate ($\sigma\sim 1$) nonlocality regimes. 

\subsection{Sound propagation}

In order to acquire understanding of shock wave behaviors, it is helpful to express Eq.~(\ref{NLS12}) in a hydrodynamic form, and hence treat the light field as a {\em classical fluid}. This can be done by using the Madelung transformation
$u(\xi,\zeta)=\sqrt{\rho(\xi,\zeta)}e^{i\phi(\xi,\zeta)}$, which results in two Euler-like fluid equations
\begin{subequations}\label{Euler1}
\begin{eqnarray}
&&\frac{\partial \rho}{\partial \zeta}+\frac{\partial}{\partial \xi}(\rho v)=0,\label{Euler1a}\\
&&\frac{\partial v}{\partial \zeta}+\frac{\partial}{\partial \xi}\left[\frac{1}{2}v^2-{\cal G}\int d\xi' g(\xi'-\xi)\rho(\xi',\zeta) \right.\notag\\
&&\left. \quad -\frac{1}{2\sqrt{\rho}}\frac{\partial^2 \sqrt{\rho}}{\partial \xi^2} \right]\label{Euler1b},
\end{eqnarray}\end{subequations}
where $\rho=|u|^2$ is the intensity and $v=\partial\phi/\partial \xi$ is the flow velocity of the light fluid. In the weak nonlocality regime, the second equation of Eq.~(\ref{Euler1}) can be written in the form
\begin{eqnarray}\label{Euler2}
&&\frac{\partial v}{\partial \zeta}+\frac{\partial}{\partial \xi}\left[\frac{1}{2}v^2+{\cal G}\rho+{\cal R}\frac{\partial^2 \rho}{\partial \xi^2}+\frac{1}{8\rho^2}\left(\frac{\partial \rho}{\partial \xi}\right)^2
\right.\notag\\
&&\left.\quad-\frac{1}{4\rho}\frac{\partial^2 \rho}{\partial \xi^2} \right]=0,
\end{eqnarray}
where the third term in the brackets on the left-hand side of Eq.~(\ref{Euler2}) comes from the intensity diffraction (characterized by the intensity diffraction parameter ${\cal R}$); the last two terms in the bracket originate from the ``quantum'' pressure. All these terms govern the formation of oscillatory waves in the hydrodynamic approach.

For a linear propagation, the probe intensity can be expressed in the form $\rho(\xi,\zeta)=\rho_b+\delta\rho(\xi,\zeta)$, where $\rho_b$ and $\delta\rho$ ($\delta\rho\ll\rho_b$) stand for the intensities of a uniform background and a small perturbation (disturbance), respectively.
In the weak nonlocality regime, the equation for the small perturbation is governed by the linear Boussinesq equation
\begin{equation}\label{Boussinesq}
\frac{\partial^2 \delta\rho}{\partial \zeta^2}-\rho_b\cG\frac{\partial^2 \delta\rho}{\partial \xi^2}+\left(\frac{1}{4}-\rho_b{\cal R}\right)\frac{\partial^4 \delta\rho}{\partial \xi^4}=0.
\end{equation}
We consider a plane-wave solution $\delta\rho=A\,e^{i(k\xi-\omega \zeta)}+{\rm c.c.}$, where $A$ is the real amplitude and
${\rm c.c.}$ means the complex conjugate. Then we get the linear dispersion relation
\begin{equation}\label{dispersion}
\omega=k\sqrt{{\cal G}\rho_b+\left(\frac14-{\cal R}\rho_b\right)k^2}.
\end{equation}
It appears that when ${\cal R}>1/(4\rho_b)$, there always exists a critical value of $k$, i.e.
$k_{\rm cr}=[{\cal G}\rho_b/({\cal R}\rho_b-1/4)]^{1/2}$,
such that $\omega$ becomes imaginary when $k>k_{\rm cr}$, corresponding to the occurrence of modulation
instability (MI). Note that the MI is unique for the defocusing nonlocal medium and crucial for the formation of various optical patterns~\cite{Shi2020}. In the long wavelength limit ($k\rightarrow0$), the dispersion relation takes the form
\begin{equation}
\omega \approx c_s k, \qquad c_s=\sqrt{\cG \rho_b},
\end{equation}
where $c_s$ is called the {\it local sound speed} of long waves on the stationary background $\rho_b$. Then, the condition of MI of a uniform background can be written as
\begin{equation}\label{ccr}
c_s>c_{cr}=\frac{1}{2}\sqrt{\frac{\cG}{\cR}},
\end{equation}
where $c_{cr}$ is the critical value of the sound speed. Fig.~\ref{fig1}(c) shows the dispersion relation~(\ref{dispersion}) for different values of the intensity diffraction parameter ${\cal R}$.
For the system parameters given in Sec.~\ref{section3A}, we have $\sigma=0.3$, ${\cal R}\approx0.1$, and $c_{cr}\approx3.16$ (see Table~\ref{Table1}).

When the small perturbation (disturbance) is increased, the nonlinearity should be taken into account and the equation of $\delta\rho$ can be written into the form of the Boussinesq equation, which supports nonlocal dark-solitons. A detailed consideration on how to get dark-soliton solutions based on Eq~(\ref{Euler1}) is presented in Appendix~\ref{Appendix2}. Figure~\ref{fig2}(b) shows the intensity profile $\rho$ of the soliton with different values of $\cR$ and $V_s$ (soliton velocity) as a function of $\xi$. One sees that the soliton width decreases with growth of ${\cR}$ for the same $V_s$, i.e. the increase of nonlocality leads to the narrowing of the dark soliton. This is because the probe intensity diffraction contributed by the nonlocality is negative [see Eq.~(\ref{NLS2})] and hence it gives an opposite effect against the normal positive diffraction, which results in an increasing of the soliton width. On the other hand, the soliton depth decreases with growth of $V_s$ for the same ${\cal R}$. We stress that the nonlocal dark solitons found here have much lower generation power than those reported before in other systems.

\section{Dispersive shock waves in two-dimensional Rydberg gases}\label{section4}

\subsection{Wave breaking and shock wave formation}

When the nonlinearity overwhelms the diffraction, a large and smooth perturbation can change its profile since each point of the perturbation propagates with a local sound speed ($c=\sqrt{{\cG}\rho}$), rather than with the
background sound speed ($c_s=\sqrt{{\cG}\rho_b}$). Consequently, higher-intensity parts of the profile will travel at a faster speed, leading to the wave steepening and, eventually, wave breaking followed by the formation of a
shock wave. In order to describe such an event occurring before the shock wave formation, we let ${\cal R}\rightarrow0$ and omit the quantum pressure in Eqs.~(\ref{Euler1b}) and
(\ref{Euler2}), arriving at the celebrated shallow-water-like equations for the light fluid of the probe field
\begin{subequations}\label{shallow_water}
\begin{eqnarray}
&&\frac{\partial \rho}{\partial \zeta}+\frac{\partial}{\partial \xi}(\rho v)=0,\\
&&\frac{\partial v}{\partial \zeta}+\frac{\partial}{\partial \xi}\left(\frac{1}{2}v^2+{\cal G}\rho\right)=0.
\end{eqnarray}\end{subequations}
To study these equations, it is convenient to cast them into the diagonal Riemann form
\begin{subequations}\label{Riemann_form}
\begin{eqnarray}
&& \frac{\prt r_+}{\prt\zeta}+\frac12(3r_++r_-)\frac{\prt r_+}{\prt\xi}=0,\\
&& \frac{\prt r_-}{\prt\zeta}+\frac12(r_++3r_-)\frac{\prt r_-}{\prt\xi}=0,
\end{eqnarray}\end{subequations}
where the Riemann invariants are given by
\begin{equation}
r_{\pm}=\frac{v}2\pm\sqrt{{\cal G}\rho}.
\end{equation}
(see, e.g., \cite{LL-6,kamch-2000}). As long as $r_+$ and $r_-$ are found, the light fluid intensity and the flow velocity are found by
\begin{equation}
\rho=\frac1{4\cG}(r_+-r_-)^2,\qquad v=r_++r_-.
\end{equation}

In the case of arbitrary initial light intensity and flow velocity, $\rho(\xi,0)$ and $v(\xi,0)$, both Riemann invariants are changing with the propagation distance $\zeta$ and the corresponding solution of
Eqs.~(\ref{Riemann_form}) can be found by the Riemann method~\cite{ikn-19}. In the following, we shall confine ourselves to the so-called {\it simple waves}, in which case we are interested in only one of the left- and
right-moving parts since both parts move independently after separation, and hence only one of the Riemann invariants is a function of $\zeta$ and the other one is a constant.

Particularly, we assume $r_-=\mathrm{const}$ and the flow velocity is equal to zero in the neighboring undisturbed region. Thus we have
\begin{equation}\label{r_minus}
r_-=\frac{v}2-\sqrt{\cG\rho}=-\sqrt{\cG\rho_b}.
\end{equation}
Then the flow velocity can be written as
$v=2\sqrt{\cG}(\sqrt{\rho}-\sqrt{\rho_b})$, and the first equation of Eqs.~(\ref{Riemann_form}) is reduced to the Hopf equation
\begin{equation}\label{Hopf}
\frac{\prt r_+}{\prt\zeta}+\frac12(3r_+-\sqrt{\cG\rho_b})\frac{\prt r_+}{\prt\xi}=0,
\end{equation}
which admits the solution (see, e.g., \cite{whitham-74,LL-6,kamch-2000})
\begin{equation}\label{Hopf_solution}
\xi-\frac12(3r_+-\sqrt{\cG\rho_b})\zeta=\overline{\xi}(r_{+,0}),
\end{equation}
with $\overline{\xi}(r_{+,0})$ the inverse function to the initial distribution $r_{+,0}=r_+(\xi,0)$ of the Riemann invariant $r_+(\xi,\zeta)$.

To be concrete, we assume that the initial light fluid intensity and the flow velocity have the form
\begin{equation}\label{initial_cond}
\rho(\xi,0)=\rho_b+\rho_h\,e^{-\xi^2/w_h^2},\quad v(\xi,0)=0,
\end{equation}
which can be easily prepared in a real experiment. Here $\rho_h$ and $w_h$ characterize, respectively, the peak intensity and the width of a Gaussian hump added on the uniform background. From Eqs.~(\ref{Hopf_solution}) and (\ref{initial_cond}) one
can readily obtain the solution
\begin{equation}\label{Hopf_solution1}
\xi(\rho)=\sqrt{{\cal G}}(3\sqrt{\rho}-2\sqrt{\rho_b})\zeta+w_h\sqrt{\ln[\rho_h/(\rho-\rho_b)]},
\end{equation}
and hence the intensity can be expressed implicitly with the solution
\begin{equation}\label{Hopf_solution2}
\rho(\xi,\zeta)=\rho_b+\rho_h\,e^{-[\xi-\sqrt{{\cal G}}(3\sqrt{\rho}-2\sqrt{\rho_b})\zeta]^2/w_h^2}.
\end{equation}

Since the flow velocity $v$ depends on the light intensity $\rho$, the hump exhibits indeed a self-steepening in the direction of propagation, resulting in a gradient catastrophe $\partial_\xi \rho(\xi,\zeta)=-\infty$ at
a certain distance $\zeta=\zeta_{\rm wb}$. This gradient catastrophe leads to the well-known wave breaking phenomenon, followed by the shock wave formation. The distance $\zeta_{\rm wb}$ is referred to as the wave breaking
distance, which can be determined from the conditions
\begin{equation}\label{wave_breaking}
\frac{\partial \xi(\rho)}{\partial \rho}=0, \quad \frac{\partial^2 \xi(\rho)}{\partial \rho^2}=0,
\end{equation}
yielding the solution
\begin{equation}\label{wave_breaking1}
\zeta_{\rm wb}=\frac{w_h}{3\sqrt{{\cal G}}}\frac{\sqrt{\rho_s+\rho_b}}{\rho_s-\rho_b}.
\end{equation}
Here $\rho_s$ is the light fluid intensity corresponding to $\partial_\xi \rho_s \rightarrow-\infty$, which can be obtained from the equation
\begin{equation}
\ln\left(\frac{\rho_h}{\rho_s-\rho_b}\right)=\frac{\rho_s}{\rho_s+\rho_b}.
\end{equation}
Although it is hard to have analytical solutions of the above equation, one can solve it numerically.

\begin{figure}
\centering
\includegraphics[width=0.8\columnwidth]{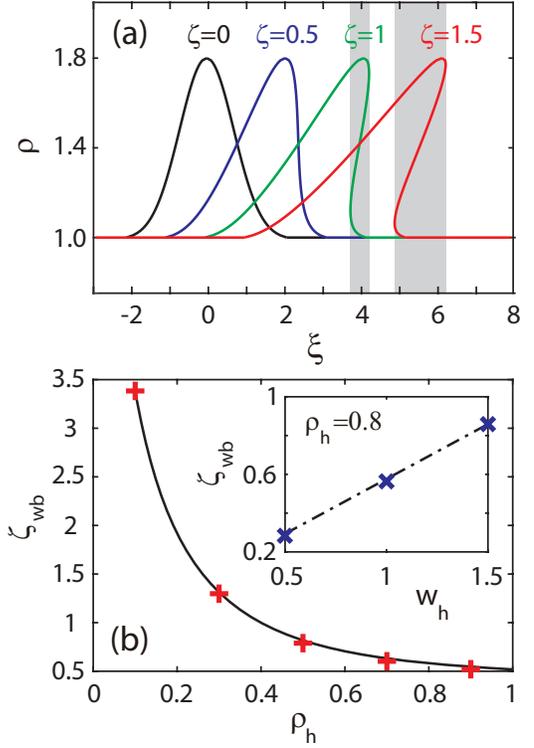}
\caption{{\footnotesize (a)~Light intensity profile $\rho$ as a function of $\xi$ for different distances $\zeta=0$, 0.5, 1, and 1.5. The hump exhibits an obvious self-steepening in the direction of propagation. A gradient catastrophe
(wave breaking) occurs at $\zeta\approx0.7$. (b)~The wave breaking distance $\zeta_{\rm wb}$ as a function of the hump's peak intensity $\rho_h$. Inset: The wave breaking distance $\zeta_{\rm wb}$ as a function of
the hump's width $w_h$. A good agreement is obtained between the analytical prediction (denoted by solid black and dash-dotted blue lines) and the numerical calculation (denoted by symbols ``+'' and ``x'', respectively).}}
\label{fig3}
\end{figure}

Figure~\ref{fig3}(a) shows the result on the light intensity profile $\rho$ as a function of $\xi$ at different propagation distance $\zeta$. We see that an obvious self-steepening of the hump occurs in the direction of propagation, which results in a wave breaking at a certain distance.
Plotted in Figure~\ref{fig3}(b) is the wave-breaking distance $\zeta_{\rm wb}$ as a function of the hump's peak intensity $\rho_h$.
In the figure, the analytical result is given by the solid black and dash-dotted blue lines, and the numerical one is denoted by the points indicated by symbols red ``+'' and blue ``x'', respectively. The dependence of $\zeta_{\rm wb}$ on the hump's width $w_h$ is illustrated in the inset of the figure.
We see that $\zeta_{\rm wb}$ decreases rapidly (slowly) with growth of $\rho_h$ when $\rho_h$ is small (large), and $\zeta_{\rm wb}$ increases linearly with growth of $w_h$. Thus, there is a good agreement between analytical predictions (denoted by lines) and numerical calculations (denoted by symbols), which means that the diffraction plays indeed no significant effect on the occurrence of the wave breaking.

Note that $\zeta_{\rm wb}$ may depend on the nonlocality degree of the Kerr nonlinearity if the system works in the strong nonlocality regime. This issue, however, has been discussed in Ref.~\cite{Ghofraniha2012} in a different setting and is outside the scope of this work.

\subsection{DSWs in the weak nonlocality regime}

After the occurrence of the wave breaking, the negligible diffraction approximation is not applicable anymore to such a problem. Therefore, we have to consider the diffraction, which can interplay with the Kerr nonlinearity. Indeed, such an interplay leads to the formation of a DSW instead of a nonphysical multi-valued solution.
In practice, one can obtain global solutions of the Whitham equations if the system under consideration belongs to the class of {\em completely integrable} equations.
If this is not true, as in our case, we have to resort to the method developed by El~\cite{el-05}, which works very well for both integrable and non-integrable equations, providing though limited but very important information about DSWs. Moreover, since the soliton velocity is always smaller than the long-wave sound speed $c_s$ [seen from Eq.~(\ref{soliton_velocity1})], the front small-amplitude edge of a DSW must propagate with a group velocity faster than that of the trailing soliton edge. El's method allows us to find the velocities of both edges of DSWs. Hereafter, we use $s_+^{\rm DSW}$ ($s_-^{\rm DSW}$) to represent the speed of the small-amplitude (soliton) edge of DSWs.

\vspace{2mm}
{\bf Small-amplitude edge of DSWs.}
In practice, it is convenient to rewrite the dispersion relation (\ref{dispersion}) in the form
\begin{equation}\label{dispersion1}
\omega=k[v+ c\alpha(c)],
\end{equation}
where $c$ denotes the local sound speed, $c=\sqrt{{\cal G}\rho}$, and
\begin{equation}\label{alpha}
\al(c)=\sqrt{1+\frac{1-c^2/c_{cr}^2}{4c^2}\,k^2},\quad k=2c\sqrt{\frac{\al^2-1}{1-c^2/c_{cr}^2}},
\end{equation}
with $c_{cr}=\sqrt{\cG/\cR}/2$ the critical value of the sound speed [defined by Eq.~(\ref{ccr})]. Following the procedure described
in~\cite{el-05,egs-06,kamch-19}, the equation for the function $\al(c)$ can be found as
\begin{equation}\label{dadc1}
\frac{d\al}{dc}=-\frac{(\al+1)[2\al+1-(2+\al)c^2/c_{cr}^2]}{(2\al+1)(1-c^2/c_{cr}^2)c}.
\end{equation}

For a right-moving DSW, its left edge corresponds to the soliton edge, which moves slowly; its right edge corresponds to the small-amplitude edge, which moves fast. Then, Eq.~(\ref{dadc1}) can be solved with the boundary condition $\al(c=c_L)=1$, where $c_L$ denotes the local sound speed at the left (soliton) edge of the DSW, given by
\begin{equation}\label{c_L}
c_L=\sqrt{\cG\rho_L},
\end{equation}
with $\rho_L$ the left-edge intensity. Such a
boundary condition means that the small-amplitude edge merges into the soliton edge where the distance between wave crests becomes infinitely large, i.e. $k\rightarrow0$.

When $\al=\al(c)$ is solved (this can be done numerically), the velocity of the
small-amplitude edge can be obtained, given by
\begin{equation}\label{small_amplitude_edge_speed}
s_+^{\rm DSW}=\left.\frac{d\om}{dk}\right|_{c=c_R}=c_R\left[2\al(c_R)-\frac1{\al(c_R)}\right],
\end{equation}
where $c_R=\sqrt{\cG\rho_R}$ denotes the local sound speed at the right (small-amplitude) edge.

\vspace{2mm}
{\bf Soliton edge of DSWs.}
The speed of the soliton edge of the DSW can be established by using the dispersion relation Eq.~(\ref{soliton_velocity1}). From the correspondence relationship Eq.~(\ref{soliton_velocity2}), we can write $\tom$ in the form
\begin{equation}\label{tom}
\tom=\tk[v+c\tal(c)],
\end{equation}
where
\begin{equation}
\tal=\sqrt{1-\frac{1-c^2/c_{cr}^2}{4c^2}\,\tk^2},\quad \tk=2c\sqrt{\frac{1-\tal^2}{1-c^2/c_{cr}^2}}.
\end{equation}
Then, the equation for the variable $\tal(c)$ is found to be
\begin{equation}\label{dadc2}
\frac{d\tal}{dc}=-\frac{(\tal+1)[2\tal+1-(2+\tal)c^2/c_{cr}^2]}{(2\tal+1)
(1-c^2/c_{cr}^2)c},
\end{equation}
which has the same form with Eq.~(\ref{dadc1}) except that in this case the equation should be solved with the boundary condition
$\tal(c=c_R)=1$. Such a boundary condition means that the soliton edge merges into the small-amplitude edge where the amplitude of oscillatory waves vanishes together with the inverse half-width of the soliton, i.e. $\tk\rightarrow0$.

The soliton velocity is given by $V_s=v+c_L\tal(c_L)$ according to Eqs.~(\ref{soliton_velocity2}) and (\ref{tom}), where the flow velocity $v$ is obtained by Eq.~(\ref{r_minus}). Therefore, the speed of the soliton
edge can be expressed as
\begin{equation}\label{soliton_edge_speed}
s_-^{\rm DSW}=V_s=c_L[2+\tal(c_L)]-2c_s.
\end{equation}
The intensity of the trailing soliton at the soliton edge can be found from Eq.~(\ref{soliton_depth1}), given by
\begin{equation}
\rho_b-\rho_m=(c_s^2-V_s^2)/\cG,
\end{equation}
where we have used the relation $\rho_b=c_s^2/\cG$.

\vspace{2mm}
{\bf Singular behavior of the DSW edge speeds.}
In the above discussion, we have found the analytical expressions of DSW edge speeds, given by Eqs. (\ref{small_amplitude_edge_speed}) and (\ref{soliton_edge_speed}). In fact, the edge speeds can be greatly increased with growth of the local sound speed, and they may exhibit a singular behavior when the local sound speed is in the vicinity of the critical value $c_{cr}$, which is dependent on the nonlocality degree $\sigma$ of the Kerr nonlinearity.

\begin{figure}
\centering
\includegraphics[width=0.8\columnwidth]{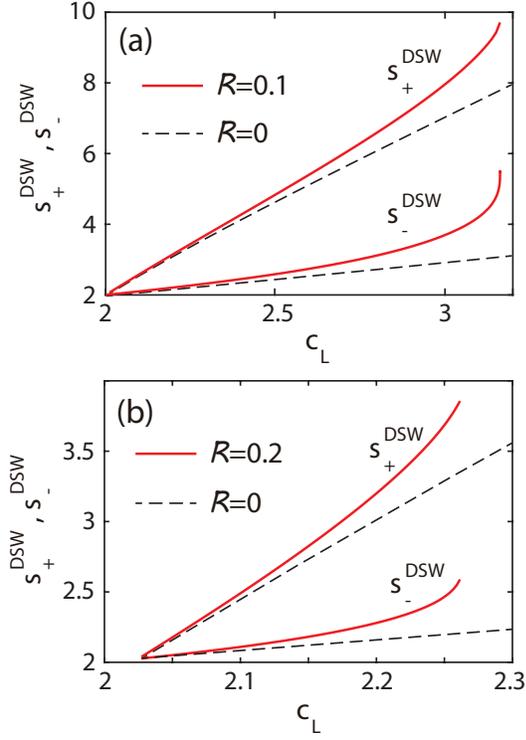}
\caption{{\footnotesize
The velocities of the small-amplitude ($s_+^{\rm DSW}$) and soliton ($s_-^{\rm DSW}$) edges, as functions of the left-edge local sound speed $c_L$ for different intensity diffraction parameter ${\cal R}$.
(a) The upper (lower) part is for $s_+^{\rm DSW}$ ($s_-^{\rm DSW}$); solid red (dashed black) lines are for ${\cal R}=0.1$ (${\cal R}=0$).
(b) The same as (a) but with ${\cal R}=0.2$ and ${\cal R}=0$.
Note that these edge speeds become singular at $c_L=c_{cr}\approx3.162$ for ${\cal R}=0.1$ [panel (a)], and at $c_L=c_{cr}\approx2.236$ for ${\cal R}=0.2$ [panel (b)]; no singular behavior occurs for the case of ${\cal R}=0$ (local medium limit).
}}
\label{fig4}
\end{figure}

To demonstrate this, we solve Eqs.~(\ref{dadc1}) and (\ref{dadc2}) with different values of the intensity diffraction parameter ${\cal R}$
in the range of $c_s<c_L<c_{cr}$  [equivalent to the range of $\rho_b<\rho_L<\rho_{cr}=1/(4{\cal R})$].
Plotted in Fig.~\ref{fig4} are the velocities of the small-amplitude and soliton edges, i.e. $s_+^{\rm DSW}$ and $s_-^{\rm DSW}$, as functions of the left-edge local sound speed $c_L$ for different intensity diffraction parameter ${\cal R}$.
%
%
From the results illustrated in the figure, we see that both $s_+^{\rm DSW}$ and $s_-^{\rm DSW}$ are increasing functions of $c_L$.
Particularly, $s_+^{\rm DSW}$ and $s_-^{\rm DSW}$ increase more and more rapidly with growth of $c_L$, and a singular behavior occurs at the critical point $c_L=c_{cr}$ for both edge speeds, corresponding to a change of the sign of the coefficient before $k^2$ in Eq.~(\ref{alpha}). Particularly, the singularity occurs at $c_L=c_{cr}\approx3.162$ ($c_L=c_{cr}\approx2.236$) for ${\cal R}=0.1$ (${\cal R}=0.2$).
However, in the case of vanishing nonlocality 
(i.e. ${\cal R}=0$), both $s_+^{\rm DSW}$ and $s_-^{\rm DSW}$ are linearly increasing functions of $c_L$, displaying no singular behavior. 
Generally, $s_+^{\rm DSW}$ and $s_-^{\rm DSW}$ for ${\cal R}>0$ are larger than those for ${\cal R}=0$ in the weak nonlocality regime.


\subsection{Numerical simulations and stability of DSWs}

The validity of the analytical results has been confirmed by carrying out numerical simulations on Eq.~(\ref{NLS2}) by taking different values of the intensity diffraction parameter ${\cal R}$. Strictly speaking, the above formulas are correct for the description of propagation of the initial step-like discontinuity which is of the simple-wave type. However, an initial pulse usually splits to two simple waves propagating in opposite directions, so these formulas provide asymptotic values of the edge speeds in a quite general situation (see more details in Ref.~\cite{kamch-19}). To be concrete, in our simulations, the initial condition is chosen as Eq.~(\ref{initial_cond}), with $\rho_b=1$, $\rho_h=2$, and $w_h=1$. 

\begin{figure*}
\centering
\includegraphics[width=1.95\columnwidth]{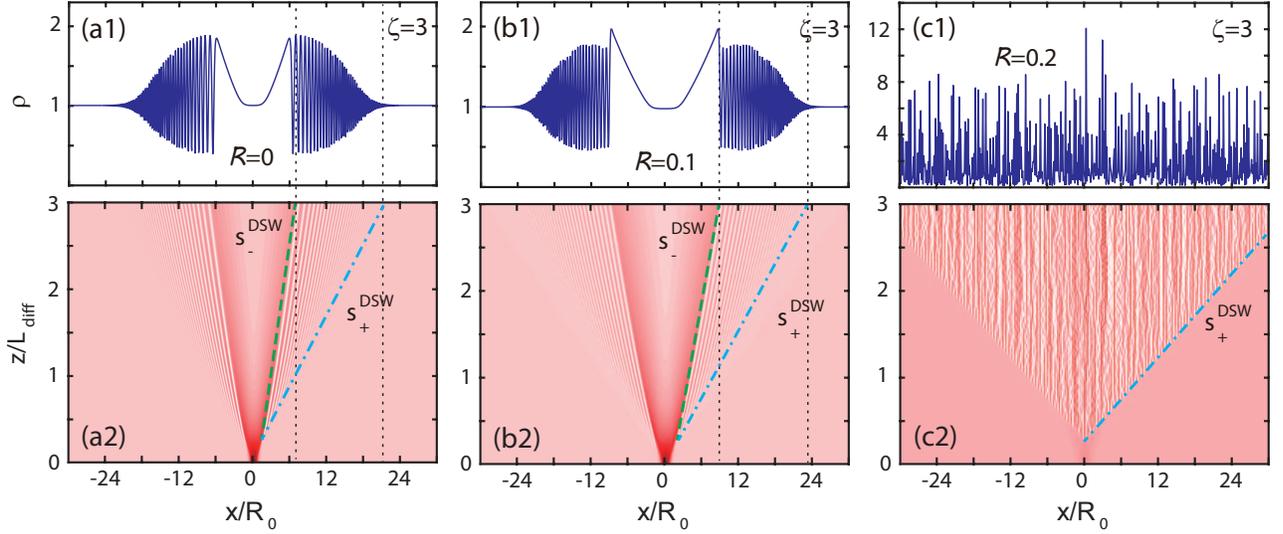}
\caption{{\footnotesize Numerical results of the formation and propagation of a DSW based on the model~(\ref{NLS2}). Snapshots of the intensity profile $\rho=|u|^2$ at $\zeta\equiv z/L_{\rm diff}=3$ (upper panels) and corresponding level plots of the propagation from $\zeta=0$ to 3 (lower panels), for the intensity diffraction parameter ${\cal R}=0$ [(a1) and (a2)], 0.1 [(b1) and (b2)], and 0.2 [(c1) and (c2)], respectively. In all columns, the initial condition is given by Eq.~(\ref{initial_cond}), with $\rho_b=1$, $\rho_h=2$ and $w_h=1$.  The slopes of dash-dotted blue (dashed green) lines in the lower panels denote the speeds of the soliton  (small-amplitude) edges of the DSW. The vertical dotted black lines in both the lower and upper panels give the positions of the soliton and small-amplitude edges of the DSW along the transverse ($x$) direction. The DSW is stable in the first (${\cal R}=0$) and second (${\cal R}=0.1$) columns, for $c_L\approx2.7<c_{cr}\rightarrow\infty$ and $c_L\approx2.9<c_{cr}\approx3.162$, respectively. However, it is unstable in the third column (${\cal R}=0.2$) for $c_L>c_{cr}\approx2.236$. }}
\label{fig5}
\end{figure*}

Shown in Fig.~\ref{fig5}(a1) and Fig.~\ref{fig5}(a2)
%
%
are respectively the probe intensity $\rho=|u|^2$ at $\zeta=3$ and its propagation result from $\zeta=0$ to 3 for ${\cal R}=0$. Fig.~\ref{fig5}(b1) and Fig.~\ref{fig5}(b2) show the same results as in Fig.~\ref{fig5}(a1) and Fig.~\ref{fig5}(a2) but for ${\cal R}=0.1$. From the first and second columns, we see clearly that in both cases the DSWs are quite stable during propagation, which is due to the fact of $c_L<c_{cr}$. Actually, the left-edge intensities in Fig.~\ref{fig5}(a1) and Fig.~\ref{fig5}(b1) are respectively given by $\rho_L\approx1.8$ and $\rho_L\approx 2$, corresponding to the left-edge local sound speeds $c_L\approx2.7$ and $c_L\approx 2.9$. Since $c_{cr}\rightarrow\infty$ ($c_{cr}\approx3.162$) for ${\cal R}=0$ (${\cal R}=0.1$), one has $c_L<c_{cr}$ in both situations. Moreover, the increase of velocities of the small-amplitude and soliton edges with growth of ${\cal R}$ ($c_L$),
found by the analytical approach in the last subsection, is also observed in the numerical simulation [see the slope of the lines of $s_+^{\rm DSW}$ and $s_-^{\rm DSW}$ in Fig.~\ref{fig5}(a2) and (b2)].

The case with larger intensity diffraction parameter, ${\cal R}=0.2$, is
also calculated, with the results presented in Fig.~\ref{fig5}(c1) and Fig.~\ref{fig5}(c2). In this case, however, the DSW becomes unstable and the instability emerges at the {\em small-amplitude} edge of the DSW.
The reason of the instability comes from that fact that the local sound speed at the left (soliton) edge, $c_{L}$, is larger than the critical value $c_{cr}$, i.e.
\begin{equation}\label{ccr1}
c_L>c_{cr}=\frac{1}{2}\sqrt{\frac{\cG}{\cR}}.
\end{equation}
Therefore, such an instability is stemmed from the MI of sound waves. 
Since solitons are usually rather stable due to the nonlocality of the Kerr nonlinearity~\cite{Bang2002}, the instability does not emerge at the soliton edge. 

\begin{figure}
\centering
\includegraphics[width=0.8\columnwidth]{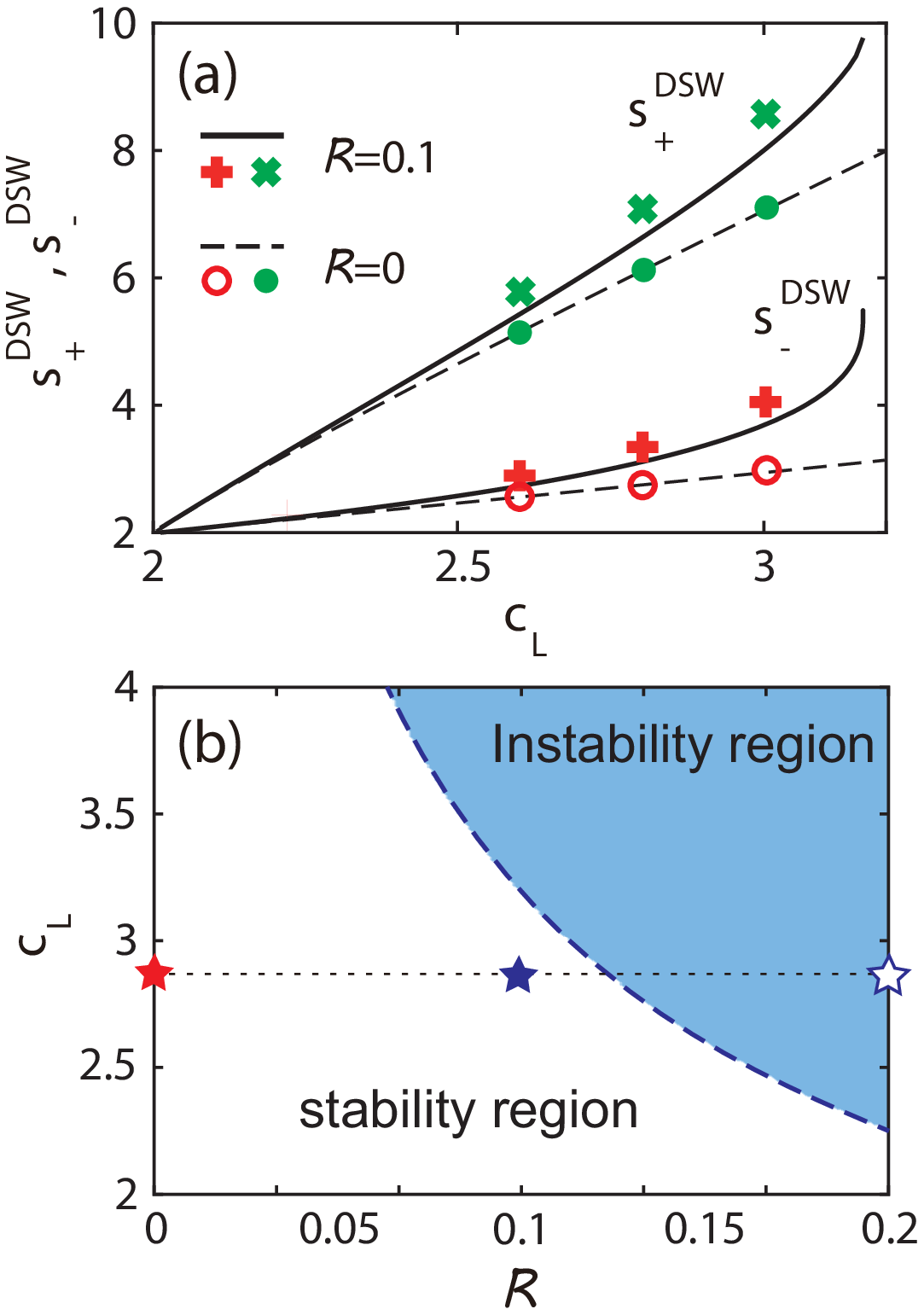}
\caption{{\footnotesize (a) Comparison between the analytical and numerical results of the small-amplitude ($s_+^{\rm DSW}$) and soliton ($s_-^{\rm DSW}$) edge speeds
as functions of the left-edge local sound speed $c_L$.
Lines are analytical results for ${\cal R}=0.1$ (solid black lines) and 0 (dashed black lines); symbols are  numerical results for ${\cal R}=0.1$ (``+'' for $s_-^{\rm DSW}$ and ``x'' for $s_+^{\rm DSW}$) and 0 (hollow circles for $s_-^{\rm DSW}$ and solid circles for $s_+^{\rm DSW}$).
(b) Stability/instability diagram of the DSW in the plane of $c_L$ and ${\cal R}$. The left-lower (white) and right-upper (cyan) domains denote respectively the stability and instability regions of the DSW. The
solid red, solid blue, and hollow blue stars correspond to the values of the DSW shown in Fig.~\ref{fig5}(a1)-(a2), (b1)-(b2), and (c1)-(c2), respectively. }}
\label{fig6}
\end{figure}

In Fig.~\ref{fig6}(a), we provide a comparison between the analytical predictions and numerical calculations of the small-amplitude ($s_+^{\rm DSW}$) and soliton ( $s_-^{\rm DSW}$) edge speeds, which are taken as functions of the left-edge local sound speed $c_L$.
%
%
A good agreement is achieved between the analytical and numerical results, implying the effectiveness of our theoretical analysis presented in the last subsection. Fig.~\ref{fig6}(b) shows the stability/instability diagram of DSWs in the plane of $(c_L, {\cal R})$. It is seen that the instability of DSWs occurs in the region where both $c_L$ and ${\cal R}$ have large values [denoted by the right-upper (cyan) domain]. Otherwise, if $c_L$ and/or ${\cal R}$ have small values, DSWs are stable. This tells us that it is possible to control the stability/instability of DSWs in an active way by changing either $c_L$ (or the left-edge intensity $\rho_L$) and ${\cal R}$ (or the nonlocality degree $\sigma$) in the present Rydberg-EIT system.

\subsection{DSWs for a moderate degree of nonlocality}

When the nonlocality degree of the Kerr nonlinearity is increased so that $\sigma\sim 1$, the system works in the intermediate nonlocality regime. In this situation, the reduced model (\ref{NLS2}) is not applicable and we have to solve the original model~(\ref{NLS12}) by using numerical methods.

\begin{figure}
\centering
\includegraphics[width=0.8\columnwidth]{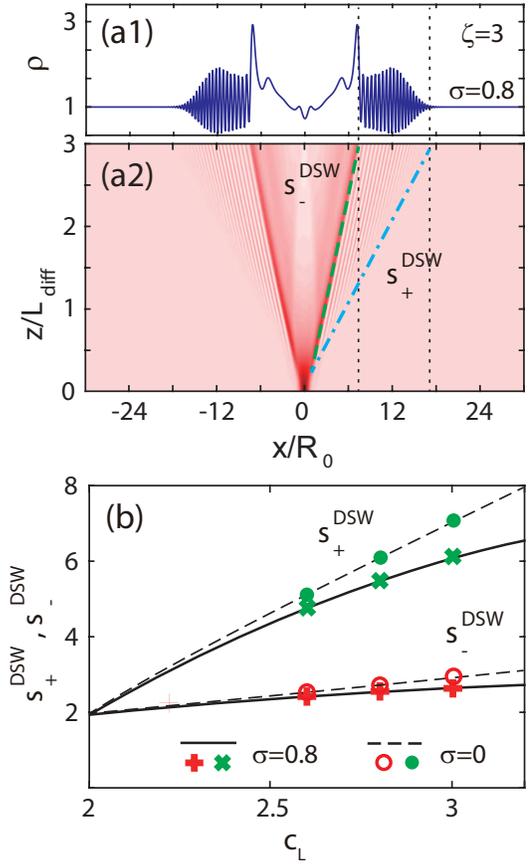}
\caption{{\footnotesize Formation and propagation of DSWs in the intermediate nonlocality regime, obtained by solving the model~(\ref{NLS1}). (a1) Snapshot of intensity profile $\rho=|u|^2$ at $\zeta=z/L_{\rm diff}=3$ for the nonlocality degree $\sigma=0.8$.
(a2) Level plot for the propagation of $\rho$ from $\zeta=0$ to 3 for $\sigma=0.8$.
(b) Velocities of the small-amplitude and soliton edges, $s_+^{DSW}$ and $s_-^{\rm DSW}$, as functions of the left-edge local sound speed $c_L$ for $\sigma=0.8$ and 0.
Dashed lines: analytical results for $\sigma=0$; solid lines: fitting curves for $\sigma=0.8$.
Symbols ``+'' and ``x'': numerical results for $\sigma=0.8$ (``+'' for $s_-^{\rm DSW}$ and ``x'' for $s_+^{DSW}$), and 0 (hollow circles for $s_-^{\rm DSW}$ and solid circles for  $s_+^{DSW}$).
}}
\label{fig7}
\end{figure}

Shown in Fig.~\ref{fig7}(a1) and Fig.~\ref{fig7}(a2) are the probe-field intensity $\rho$ at $\zeta=3$ and its propagation from $\zeta=0$ to 3 for $\sigma=0.8$, respectively.
%
%
The initial condition is the same with that used in Fig.~\ref{fig5}. As one can see, the left-edge local sound speed $c_L$, and hence the left-edge intensity $\rho_L$, is increased significantly due to growth of the nonlocality degree $\sigma$. To be specific, one has $c_L\approx3.5$ ($\rho_L\approx3.0$) for $\sigma=0.8$. Fig.~\ref{fig7}(b) shows the velocity of the small-amplitude edge, $s_+^{\rm DSW}$, and the one of the soliton edge, $s_-^{\rm DSW}$, as functions of the left-edge local sound speed $c_L$ for $\sigma=0.8$. In contrast with the DSW edge speeds in the weak nonlocality regime, they increase more and more slowly with growth of $c_L$ and are slower than those for $\sigma=0$ in the intermediate nonlocality regime. Moreover, the DSW edge speeds show no singular behavior in this regime and no instability occurs, implying the increment of nonlocality can suppress the instability of DSWs.

\section{Dispersive shock waves in three-dimensional Rydberg gases}\label{section5}

We now turn to the investigation on DSWs in a 3D Rydberg gas. Because in this situation the analytical approach employed in the last section is not applicable, we have to resort to numerical simulations. It is well known that high-dimensional localized nonlinear excitations are usually unstable in Kerr media, thus the suppression of such instability is one of the great challenges. Nevertheless, here we show that the instability of DSWs in a 3D Rydberg gas can be arrested by the giant nonlocal Kerr nonlinearity contributed by the Rydberg-Rydberg interaction; the active control over DSWs can also be effectively realized by using the Rydberg-EIT system.

\subsection{DSWs for the case of weak dispersion (${\cal D}=0$)}

We look for DSWs in the form
\begin{equation}
U(\zeta,\tau,\xi,\eta)=p(\zeta,\tau)\,u(\zeta,\xi,\eta),
\end{equation}
where the wavepacket $p(\zeta,\tau)$ is still given by Eq.~(\ref{wave_packet}), but $u$ is governed by the (3+1)D wave equation
\begin{align}\label{NLS3}
&i \frac {\partial u}{\partial \zeta}+{\cal D}\frac{\partial^2 u}{\partial \tau^2}+\frac{1}{2}\left(\frac{\partial^2}{\partial \xi^2}+\frac{\partial^2}{\partial \eta^2}\right) u \notag\\
&\quad + {\cal G}\iint d\xi' d\eta' g(\xi'-\xi,\eta'-\eta)|u(\xi',\eta',\zeta)|^2\,u=0,
\end{align}
with $\zeta$, $\tau$, $\xi$, and $\eta$ as independent variables. Here, $\eta=y/R_0$ is the other transverse coordinate; the nonlocal response function
$g(\xi'-\xi,\eta'-\eta)=(1/G_0)\int d\zeta\, G(\xi'-\xi,\eta'-\eta,\zeta)$, obeying $\iint d\xi d\eta \,|g(\xi,\eta)|=1$; and other parameters in Eq.~(\ref{NLS3}) are the same as those used in Eq.~(\ref{NLS1}). For the parameters of cold $^{87}$Rb atoms, the expression of the nonlocal response function in Eq.~(\ref{NLS3}) can be simplified as the form
\begin{eqnarray}\label{response_2D}
&& g(\Delta\xi,\Delta\eta) \approx -g_0\int d\zeta \left\{g_1+\frac{g_2}{\sigma^6}\left[\Delta\xi^2+\Delta\eta^2+\frac{z^2}{R_0^2}\right]^3\right\}^{-1},
\end{eqnarray}
with $\Delta\xi=\xi'-\xi$, $\Delta\eta=\eta'-\eta$, and $g_{0,1,2}$ being the same with those used in Eq.~(\ref{response}).

Following the line of the above two sections, under the condition of negligible group-velocity dispersion (i.e. ${\cal D}\approx 0$; see the discussion given in Sec.~\ref{section3A}) and in the weak nonlocality regime (i.e. $\sigma\ll1$), Eq.~(\ref{NLS3}) can be reduced to a simple model:
$i \partial u/\partial \zeta+(1/2)(\partial^2/\partial \xi^2+\partial^2/\partial \eta^2)u - {\cal G}|u|^2\,u - [{\cal R}_x\partial^2 |u|^2/\partial \xi^2
+{\cal R}_y\partial^2 |u|^2/\partial \eta^2]u =0$,
where ${\cal R}_x$=$-({\cal G}/2)\int \xi^2 g(\xi,\eta) \,d\xi$ and ${\cal R}_y$=$-({\cal G}/2)\int \eta^2 g(\xi,\eta) \,d\eta$ are intensity diffraction parameters in the $x$ and $y$ directions, respectively. However, we find that such a reduced model can not support stable DSWs for any values of ${\cal R}_x$ and ${\cal R}_y$, i.e. the weak nonlocality of the Kerr nonlinearity cannot prevent the occurrence of instability in two transverse directions.

\begin{figure*}
\centering
\includegraphics[width=1.95\columnwidth]{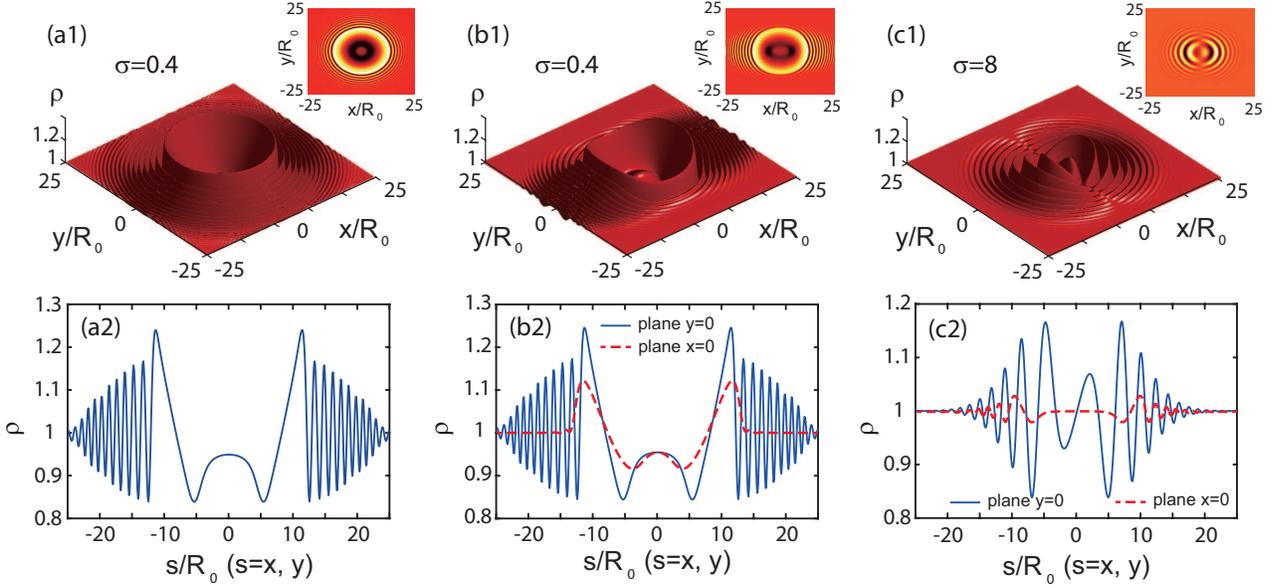}
\caption{{\footnotesize DSWs obtained by numerically solving NNLSE (\ref{NLS3}) in the intermediate nonlocality regime
for dimensionless dispersion parameter ${\cal D}\approx 0$.
(a1) and (a2) Surface plot and profile of the probe-field intensity of an {\it isotropic DSW} for the nonlocality degree $\sigma=0.4$, by taking
$\rho=|u|^2$ as a function of $x/R_0$ and $y/R_0$ at $z=4L_{\rm diff}$.
Inset: level plot of the isotropic DSW. The initial condition is given by Eq.~(\ref{initial_condition_2D1}), with $\rho_b=\rho_h=1$ and $w_{h,x}=w_{h,y}=2$.
(b1) and (b2) Surface plot and profile of the probe-field intensity of a {\it type-I anisotropic DSW} for $\sigma=0.4$. The initial condition is the same as that used in (a1) and (a2) but with $w_{h,y}=2\sqrt{10}$.
(c1) and (c2) Surface plot and profile of the probe-field intensity of a {\it Type-II  anisotropic DSW} for $\sigma=8$. The initial condition is given by Eq.~(\ref{initial_condition_2D2}), with $\rho_b=\rho_h=1$ and $w_{h,x}=w_{h,y}=1$.
In (b2) and (c2), the solid (dashed) line illustrates $\rho$ on the cross-sectional plane $y=0$ ($x=0$).}}
\label{fig8}
\end{figure*}

This fact tells us that, to suppress the occurrence of the instability of DSWs in 3D gas, one must increase the nonlocality degree $\sigma$~\cite{Armaroli2009}, and hence the system must work in the intermediate nonlocality regime ($\sigma \sim 1$) and we need to solve Eq.~(\ref{NLS3}) instead of the reduced model. Fig.~\ref{fig8} shows the results on the formation and propagation of DSWs in 3D gas for $\sigma=0.4$, obtained by numerically solving Eq.~(\ref{NLS3}) under the condition of negligible group-velocity dispersion (${\cal D}\approx 0$).
%
%
Compared with DSWs in 2D gas, DSWs in 3D gases allow diverse profiles of the probe-field intensity and exhibit richer physical phenomena.

Panels (a1) and (a2) of Fig.~\ref{fig8} present the surface plot of an {\it isotropic DSW}, by taking the probe-field intensity $\rho=|u|^2$ as a function of $x/R_0$ and $y/R_0$ at $\zeta=z/L_{\rm diff}=4$ (its level plot is shown in the inset on the upper right) and the corresponding profile on the cross-sectional plane $y=0$ ($x=0$), respectively. When implementing the calculation, we have used the transformation
$u(\xi,\eta,\zeta)=\sqrt{\rho(\xi,\eta,\zeta)}e^{i\phi(\xi,\eta,\zeta)}$ and the following initial condition
\begin{eqnarray}\label{initial_condition_2D1}
&& \rho(\xi,\eta,0)=\rho_b+\rho_h\,e^{-\xi^2/w_{h,x}^2-\eta^2/w_{h,y}^2},\notag\\
&& v(\xi,\eta,0)=0.
\end{eqnarray}
Here $\rho_b$ and $\rho_h$ are respectively the uniform background and the Gaussian peak intensity of the probe field; $w_{h,x}$ and $w_{h,y}$ are respectively hump's widths along the $x$ and $y$ directions. In the example, we have taken $\rho_b=\rho_h=1$ and $w_{h,x}=w_{h,y}=2$. In this case, the DSW obtained is isotropic in both transverse (i.e. $x$ and $y$) directions and it is quite stable during propagation.

The system also supports other kinds of DSWs, which can be obtained
by considering $w_{h,x}\neq w_{h,y}$. Panels (b1) and (b2) show respectively the surface plot and profile of the DSW for $\sigma=0.4$, with the initial condition the same as that used in (a1) and (a2) except that $w_{h,y}=2\sqrt{10}$;  the solid (dashed) line in panel (b2) illustrates $\rho$ on the cross-sectional plane $y=0$ ($x=0$). We see that, since $w_{h,x}<w_{h,y}$, the wave breaking occurs firstly in the $x$ direction, where a fast oscillation structure appears;
in this case, the DSW obtained has symmetric intensity profile in $x$ or $y$ direction, but they are different from each other. We call such a DSW as the {\it type-I anisotropic DSW} for convenience.

A different type of anisotropic DSWs from the one given in panels (b1) and (b2) can also be found. Illustrated in panels (c1) and (c2) is an anisotropic DSW for $\sigma=8$, obtained by using the following initial condition
\begin{eqnarray}\label{initial_condition_2D2}
&& \rho(\xi,\eta,0)=\rho_b-\rho_h\,\xi\,e^{-\xi^2/w_{h,x}^2-\eta^2/w_{h,y}^2},\notag\\
&& v(\xi,\eta,0)=0,
\end{eqnarray}
where $\rho_h=1$ and $w_{h,x}=w_{h,y}=2$. From the figure, we see that
the intensity of the DSW is symmetric in one transverse direction and anti-symmetric in the other transverse direction~\cite{Marcucci2020}. We call such a DSW as the {\it type-II anisotropic DSW}. 

\subsection{DSWs for the case of large dispersion (${\cal D}= 0.5$)}

In the above discussion, the group-velocity dispersion of the system has been disregarded, which is valid only for cases where the time duration of the probe field $\tau_0$ is large enough (and hence the dimensionless parameter ${\cal D}\approx 0$). If $\tau_0$ is shortened so that the dispersion length $L_{\rm disp}$ of the system is decreased, the group-velocity dispersion effect of the system will play a significant role for the formation and propagation of DSWs. For example, when $\tau_0=0.33\,\mu$s, one has $L_{\rm disp}\approx0.94$\,mm, and hence ${\cal D}=0.5$. In such a situation, the terms of the dispersion and diffraction
in Eq.~(\ref{NLS}) must be treated at the same footing, and the dimensionless half Rabi frequency of the probe field cannot be factorized anymore. Then,
the dimensionless form of the (3+1)D  NNLSE (\ref{NLS}) can be written as the form
\begin{align}\label{NLS4}
&i\frac {\partial U}{\partial \zeta}+\frac{1}{2}\left(\frac{\partial^2 }{\partial T^2}
+\frac{\partial^2 }{\partial \xi^2}+\frac{\partial^2 }{\partial \eta^2}\right)U  \notag\\
&+ {\cal G}\iint d\xi' d\eta' g(\xi'-\xi,\eta'-\eta)|U(\xi',\eta',\zeta,T)|^2U=0,
\end{align}
where $T=\tau-\zeta/\lambda$, with definitions of other dimensionless quantities being the same as those given in Eq.~(\ref{NLS3}).

\begin{figure}
\centering
\includegraphics[width=0.95\columnwidth]{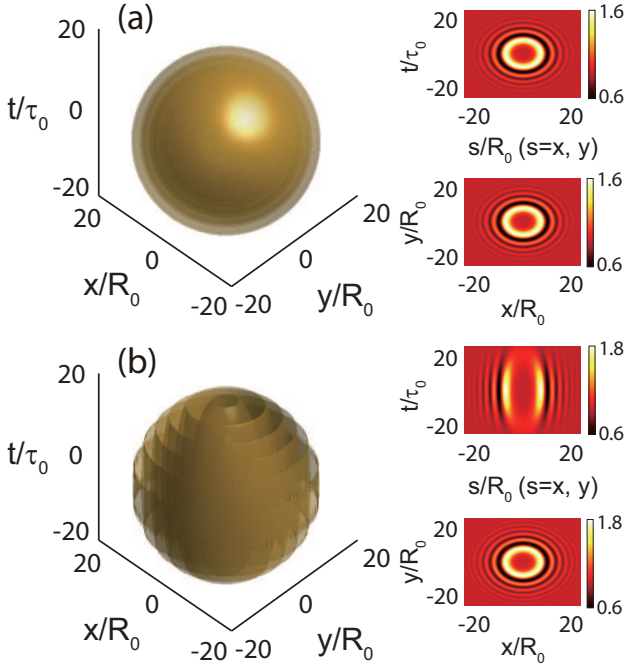}
\caption{{\footnotesize Spatiotemporal DSWs obtained by solving (3+1)D equation (\ref{NLS4}) in the intermediate nonlocality regime ($\sigma=0.4$) for dimensionless dispersion parameter ${\cal D}= 0.5$.
(a)~Surface plot of the probe-field intensity of an {\it isotropic DSW} at $z=3L_{\rm diff}$, by taking $\rho=|U|^2$ as a function of $x/R_0$, $y/R_0$, and $t/\tau_0$.
Inset: level plots of the isotropic DSW on the cross-sectional planes $y=0$ (upper part) and $t=0$ (lower part), respectively. The initial condition used is given by Eq.~(\ref{initial_condition_3D}), with $\rho_b=\rho_h=1$ and $w_{h,x}=w_{h,y}=w_{h,t}=2$.
(b)~Similar to (a) but for an {\it anisotropic DSW}. The initial condition is the same as that used in (a) but with $w_{h,t}=2\sqrt{10}$. }}
\label{fig9}
\end{figure}

Shown in Fig.~\ref{fig9}
%
%
is the result of a DSW obtained by numerically solving (3+1)D equation (\ref{NLS4}) in the intermediate nonlocality regime, $\sigma=0.4$. Since the wave breaking and the oscillatory structure appear in both two transversal spatial dimensions and the time dimension, such DSW is indeed a spatiotemporal one. Panel (a) shows the surface plot of the probe-field intensity of an {\em isotropic spatiotemporal DSW} at $z=3L_{\rm diff}$, by taking $\rho=|U|^2$ as a function of $x/R_0$, $y/R_0$, and $t/\tau_0$. The insets present level plots of this isotropic DSW on the cross-sectional planes $y=0$ (upper part) and $t=0$ (lower part), respectively. The initial condition is given by
\begin{eqnarray}\label{initial_condition_3D}
&& \rho(\xi,\eta,\tau,0)=\rho_b+\rho_h\,e^{-\xi^2/w_{h,x}^2-\eta^2/w_{h,y}^2
-\tau^2/w_{h,t}^2},\notag\\
&& v(\xi,\eta,\tau,0)=0,
\end{eqnarray}
where $w_{h,j}$ ($j=x,y$) and $w_{h,t}$ denote, respectively, the pulse's spatial and time widths, with other parameters the same as those used in Eq.~(\ref{initial_condition_2D1}). In the numerical calculation, we have taken $\rho_b=\rho_h=1$ and $w_{h,x}=w_{h,y}=w_{h,t}=2$.

Shown in Fig.~\ref{fig9}(b) is the surface plot of the probe-field intensity of an DSW by taking the initial condition basically the same as that used in Fig.~\ref{fig9}(a), but with $w_{h,t}=2\sqrt{10}$. In this case, the intensity profile of the DSW is not isotropic and the result obtained is an {\em anisotropic spatiotemporal DSW}. It should be stressed that the spatiotemporal DSW found here are quite stable during propagation, which is due to the contribution of the giant nonlocal Kerr nonlinearity that can balance the effects of the dispersion and diffraction in the system.

\begin{figure}
\centering
\includegraphics[width=0.93\columnwidth]{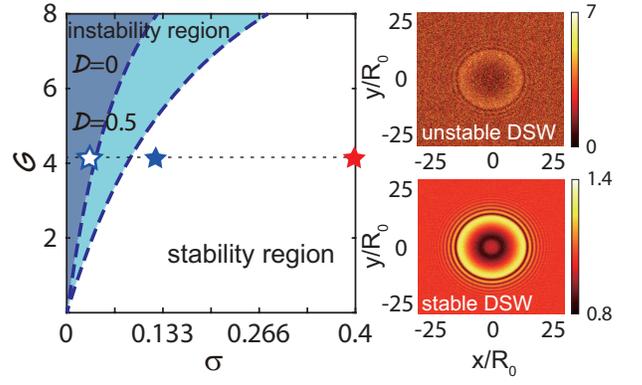}
\caption{{\footnotesize Stability diagram of DSWs in the 3D gas in the plane of the nonlocality degree $\sigma$ and the nonlinearity strength ${\cal G}$. The lower-right (white) and upper-left (cyan and dark cyan) domains denote the stability and instability regions of DSWs, respectively.
The dark cyan domain is the instability region of the DSW for the case of weak dispersion (i.e. ${\cal D}= 0$); both the cyan and dark cyan domains consist of the instability region of the DSW for the case of strong dispersion (i.e. ${\cal D}= 0.5$). The hollow blue star in the instability region corresponds to the value of $(\sigma,{\cal G})=(0.03,4.1)$; the solid blue and solid red stars in the stability region give the values of $(\sigma,{\cal G})=(0.1,4.1)$ and those for DSWs shown in Fig.~\ref{fig8}(a1)-(b2) and Fig.~\ref{fig9}, respectively. Inset: level plots of the unstable DSW (upper part) and the stable DSW (lower part), corresponding to the values for hollow blue and solid blue stars, respectively. }}
\label{fig10}
\end{figure}

The stability diagram of DSWs in the 3D Rydberg gas is illustrated in Fig.~\ref{fig10},
%
%
plotted in the plane of the nonlocality degree $\sigma$ and the nonlinearity strength ${\cal G}$. In the figure, the lower-right (white) and upper-left (cyan plus dark cyan) domains in the figure represent the stability and instability regions of DSWs, respectively. The dark cyan domain denotes the instability region of DSWs for the case of negligible dispersion (i.e. ${\cal D}\approx 0$); both the cyan and dark cyan domains consist of the instability region of DSWs for the case of strong dispersion (i.e. ${\cal D}= 0.5$). The hollow blue star in the instability region corresponds to the value of $(\sigma,{\cal G})=(0.03,4.1)$; the solid blue and solid red stars in the stability region give the values of $(\sigma,{\cal G})=(0.1,4.1)$ and $(0.4,4.1)$, corresponding to those for the DSWs shown in Fig.~\ref{fig8}(a1)-(b2) and Fig.~\ref{fig9}. From the stability diagram, one can find DSWs are unstable for small $\sigma$ and large ${\cal G}$; nevertheless, they are stable for large $\sigma$ and small ${\cal G}$. Further,
the instability region of DSWs for the case of strong dispersion (${\cal D}=0.5$) is much lager than that for the case of negligible dispersion (${\cal D}\approx 0$). Thus, for a fixed value of ${\cal G}$, the DSWs with the group-velocity dispersion require a larger value of $\sigma$ for arresting instability than that required by the DSWs without the group-velocity dispersion. For example, when ${\cal G}\approx4.1$, a stable DSW in the 3D gas can be obtained if $\sigma\gtrsim0.04$ for ${\cal D}\approx 0$ and $\sigma\gtrsim0.09$ for ${\cal D}\approx 0.5$.

\subsection{Generation power of DSWs}

The generation power of DSWs described above can be estimated by computing the corresponding Poynting's vector integrated over the cross-sectional area of the probe beam, i.e., $P=\int dS (\mathbf{E}_{p}\times \mathbf{H}_{p})\cdot \mathbf{e}_z$, where $\mathbf{e}_z$ is the unit vector in the propagation direction (i.e. the $z$ direction)~\cite{Bai2019}. Assuming that the directions of the electric and magnetic fields are
along the $x$ and $y$ directions, respectively, i.e. $\mathbf{E}_{p}=(E_{p},0,0)$ and $\mathbf{H}_{p}=(0,H_{p},0)$, with the relation $H_{p}=\varepsilon_0cn_{p}E_{p}$ ($n_{p}$ is the refractive index), one can obtain
\begin{equation}\label{GPower}
P_{\mathrm{gen}}=2\varepsilon_0cn_{p}S_0|E_{p}|_{\rm max}^2=2\varepsilon_0cn_{p}S_0\left(\frac{2\hbar}{p_{13}}\right)^2|\Omega_{p}|_{\rm max}^2,
\end{equation}%
where $S_0$ denotes the cross-sectional area of the probe beam, i.e. $S_0\approx w_{h}^2$ ($\approx w_{h,x}w_{h,y}$) for DSWs in 2D (3D) gas.

Based on Eq.~(\ref{GPower}) with the system parameters of Rydberg $^{87}$Rb atomic gas, we obtain
\begin{equation}
P_{\mathrm{gen}}\approx\left\{\begin{array}{cc}
                                3.5 \, {\rm nW}, \quad {\rm DSWs \,\, in\,\, 2D\,\, gas,} \\
                                7.0 \, {\rm nW}, \quad {\rm DSWs \,\, in\,\, 3D\,\, gas.}
                              \end{array}
                       \right.
\end{equation}%
Note that in the above calculation, the power of the uniform background of light is excluded. In contrast with other systems, the generation power of DSWs in the present system is extremely weak. The physical reason is that the present system possesses giant Kerr nonlinearity attributed to the strong Rydberg-Rydberg interaction. Such low generation power is important not only for the generation of quantum
DSWs~\cite{Simmons2020} but also for the applications of DSWs in fields of optical information transmission at very-weak-light level.

\section{Storage and retrieval of DSWs}\label{section6}

Another challenging problem in the research of DSWs is how to realize their active manipulations. Here we show that DSWs found in the present Rydberg-EIT system can be actively controlled; especially, the control field in the model may be taken as a knob to realize their storage and retrieval, similar to the photonic memory realized in conventional EIT-based systems~\cite{Fleischhauer2000,Liu2001,Phillips2001,Gorshkov2007,Shucker2008,
Novikova2012,Chen2013,Maxwell2013b,Dudin2013,Heinze2013,Wu2013,Chen2014,Hsiao2018}.

In order to realize the switching-off and switching-on of the control field, we assume that the half Rabi frequency of the control field $\Omega_c$ is not a constant but a slowly-varying function of time. Particularly, for $t<T_{\rm off}$, $\Omega_c$ is switched on; at the time interval $T_{\rm on}-T_{\rm off}$, it is switched off; then, at time  $t=T_{\rm on}$ it is switched on again. For the convenience of numerical simulations, we model such a time sequence of the control field by the following function
\begin{equation}
\Omega _{c}(t)=\Omega _{c0}\left[ 1-\frac{1}{2}\tanh \left(\frac{t-T_{\mathrm{off}}}{T_{s}}\right)+\frac{1}{2}\tanh \left(\frac{t-T_{\mathrm{on}}}{T_{s}}\right)\right],
\label{Omega}
\end{equation}
where $\Omega_{c0}$ is the amplitude; $T_{\mathrm{off}}$ and $T_{\mathrm{on}}$ are respectively times of switching-off and switching-on;
$T_{s}$ is the time characterizing the switching duration. The time sequence of the control field is depicted in Fig.~\ref{fig11} (blue curves).



Due to the time-dependence of $\Omega _{c}$, some system parameters (e.g. group velocity $V_g$, group-velocity dispersion $K_2$, absorption coefficient $A$, etc.) also become slowly-varying functions of time. Table~\ref{Table2} lists the expressions of $V_g$, $K_2$, and $A$ when $\Omega _{c}$ is switched off and on under the conditions of large one-photon detuning ($|\Delta_2|\gg \gamma_{21}$) and zero two-photon detuning ($\Delta_3= 0$).
%
\begin{table*}
\renewcommand\arraystretch{1.1}
\setlength{\tabcolsep}{3mm}
\caption{Some system parameters and characteristic lengths for the switching-on and switching-off of the control field under the conditions $|\Delta_2|\gg \gamma_{21}$ and $\Delta_3=0$. For more details, see the text.}
{\begin{tabular}{cccccc} \hline
& & & \\[-6pt]
Parameters & $V_g$ & $K_2$ & $A$ & $L_{\rm disp}$ & $L_{\rm abso}$ \\ \hline
& & \\[-6pt]
Switching on ($|\Omega_c|\sim\Delta_{2}$) & $|\Omega_c|^2/\kappa_{12}$ & $2\kappa_{12}\Delta_2/|\Omega_c|^4$ & 0 & $\tau_0^2|\Omega_c|^4/(\kappa_{12}\Delta_2)$ & $\infty$ \\ \hline
Switching off ($\Omega_c\approx0$) & 0 & $-2\kappa_{12}/\Delta_2^3$ & $\kappa_{12}\Gamma_2/\Delta_2^2$ &
$\tau_0^2\Delta_2^3/\kappa_{12}$ & $\Delta_2^2/(\kappa_{12}\Gamma_2)$ \\ \hline
& & & \\[-6pt]
\end{tabular}} \label{Table2}
\end{table*}
We see that, when the control field is switched off ($\Omega_c\approx 0$), $V_g$ is decreased from $|\Omega_c|^2/\kappa_{12}$ to nearly zero (which corresponds to the slowing down and halt of the probe field in the atomic medium), $K_2$ is changed from $2\kappa_{12}\Delta_2/|\Omega_c|^4$ to $-2\kappa_{12}/\Delta_2^3$, and $A$ is increased from zero to $\kappa_{12}\Gamma_2/\Delta_2^2$ (which corresponds to a strong absorption of the probe field in the atomic medium). Based on these results, we have the following conclusions: (i)~the absorption term on the right hand side of Eq.~(\ref{NLS}) cannot be neglected due to the increase of $A$ when the control field is switched off; (ii)~the group-velocity dispersion is weak as long as the time duration $\tau_0$ of the probe field is large enough (say, $\tau_0\ge\, 5\,\mu$s) so that ${\cal D}=L_{\rm disp}/L_{\rm diff}\ll 1$ always holds (see Table~\ref{Table2}).

\subsection{Storage and retrieval of DSWs in 2D Rydberg gas}

We first consider the storage and retrieval of DSWs in a 2D Rydberg gas. The pulse solution of Eq.~(\ref{NLS1}) can be found by using the factorization
\begin{equation}
U(\zeta,\tau,\xi)=p(\zeta,\tau)q(\tau,\xi),
\end{equation}
where the wave packet $p(\zeta,\tau)$ is given by Eq.~(\ref{wave_packet}) and $q(\tau,\xi)$ is governed by the (1+1)D equation
\begin{equation}\label{NLS51}
\frac{i}{\lambda}\frac{\partial q}{\partial \tau}+\frac{1}{2}\frac{\partial^2 q}{\partial \xi^2} + {\cal G}\int d\xi' g(\xi'-\xi)|q(\xi',\tau)|^2q=-i{\cal A}q.
\end{equation}
Here, the coefficient ${\cal A}=L_{\rm diff}/L_{\rm abso}$, with $L_{\rm abso}=1/A$ the characteristic absorption length. When obtaining Eq.~(\ref{NLS51}), we have assumed that the group-velocity dispersion in the system is
negligible (i.e. ${\cal D}\approx 0$), and integrated over the variable $\zeta$. Note that, in contrast with the function $u$ in Eq.~(\ref{NLS12}), which is only space-dependent, $q$ in the above equation is time-dependent and hence describes the time evolution of the probe field.

\begin{figure}
\centering
\includegraphics[width=1\columnwidth]{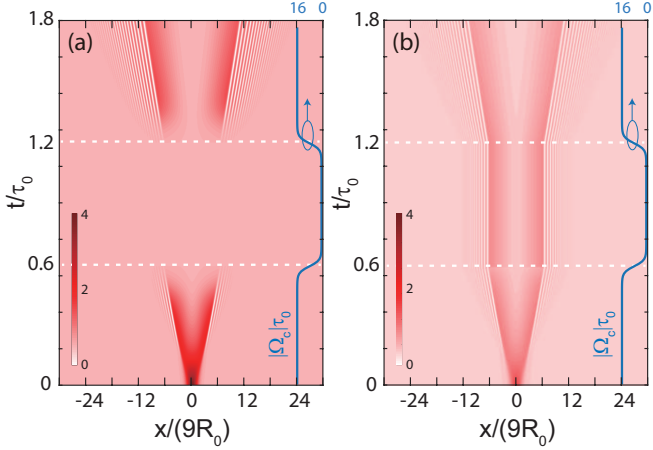}
\caption{{\footnotesize Storage and retrieval of a DSW in a 2D Rydberg gas.
(a)~Evolution of probe-field intensity $\rho=|U|^2$ as a function of $x/(9R_0)$ and $t/\tau_0$ in the course of the storage and retrieval of the DSW.
The time intervals of $0<t=0.6\tau_0$, $0.6\tau_0<t<1.2\tau_0$, and $t>1.2\tau_0$ are ones before the storage, during the storage, and after the storage (retrieval), respectively.
The corresponding $|\Omega_c|\tau_0$ as a function of $t/\tau_0$ is shown by the blue solid line on the right part of the figure. The lower and upper white dashed lines denote the times of switching-off and switching on of $\Omega_c$ (i.e. $T_{\mathrm{off}}$ and $T_{\mathrm{on}}$), respectively.
(b)~The same as (a) but for the matrix element $\rho_{13}\equiv \langle \hat{S}_{13}\rangle$.
}}
\label{fig11}
\end{figure}

To simulate the whole process of storage and retrieval of DSWs, we must solve the original model, i.e. the MB equations (\ref{Max}) and (\ref{Bloch0}), numerically. Shown in Fig.~\ref{fig11}(a)
%
%
is the spatial-temporal evolution of probe-field intensity $\rho=|U|^2$ as a function of $x$ and $t$ in the course of the storage and retrieval of a DSW. From the figure, we see that the DSW propagates in the gas when $t<0.6\tau_0$ where the control field $\Omega_c$ is switched on. Then, the DSW disappears in the time interval of $0.6\tau_0<t<1.2\tau_0$ where $\Omega_c$ is switched off, which means that the DSW is stored in the atomic medium in this time interval. Lastly, it reappears at $t>1.2\tau_0$, which means that the DSW is retrieved when the control field is switched on again. The retrieved profile has nearly the same shape as the one before the storage, except a slight attenuation due to the weak dissipation contributed by the spontaneous emission and dephasing in the system.

Drawn in panel (b) is the same as panel (a) but for the matrix element $\rho_{13}\equiv \langle \hat{S}_{13}\rangle$ (called atomic spin wave). One sees that, before $t=0.6\tau_0$ and after $t=\tau_0$, $\rho_{13}$ has a similar wave shape with $|U|^2$; however, it becomes a constant in the time interval of $0.6\tau_0<t<1.2\tau_0$. Since the probe field is stored in the form of spin wave $\rho_{13}$ when the control field is switched off and is retained until the control field is switched on again, $\rho_{13}$ can be taken as the intermediary for the storage and retrieval of the probe DSW.


\subsection{Storage and retrieval of DSWs in 3D Rydberg gas}

For a 3D Rydberg gas with negligible group-velocity dispersion (${\cal D}=0$), the solution of Eq.~(\ref{NLS1}) can be found with the form
\begin{equation}\label{Factorization}
U(\zeta,\tau,\xi,\eta)=p(\zeta,\tau)q(\tau,\xi,\eta),
\end{equation}
where the function $q(\tau,\xi,\eta)$ is governed by the (2+1)D equation
\begin{align}\label{NLS52}
&\frac{i}{\lambda}\frac{\partial q}{\partial \tau}+\frac{1}{2}\left(\frac{\partial^2}{\partial \xi^2}+\frac{\partial^2}{\partial \eta^2}\right)q + {\cal G}\iint d\xi' d\eta' g_{2D}(\xi'-\xi,\eta'-\eta) \notag\\
&\quad \times |q(\xi',\eta',\tau)|^2q=-i{\cal A}q.
\end{align}
For a 3D gas with non-negligible group-velocity dispersion (i.e. ${\cal D}\ne 0$), one cannot factorize $U$ similar to (\ref{Factorization}) anymore. In this case the evolution of $U$ is governed by an equation similar to Eq.~(\ref{NLS4}), where the absorption term $-i{\cal A}U$ should be added on the right hand side of the equation. 

\begin{figure}
\centering
\includegraphics[width=1\columnwidth]{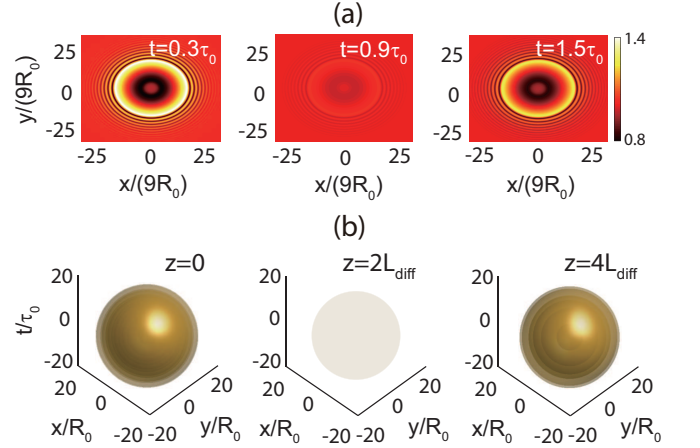}
\caption{{\footnotesize Storage and retrieval of DSWs in a 3D Rydberg gas.
(a)~Level plots describing the evolution of the probe-field intensity in the course of the storage and retrieval of a DSW  with negligible group-velocity dispersion (${\cal D}\approx0$), by taking $\rho=|q|^2$ as a function of $x/(9R_0)$ and $y/(9R_0)$, at times $t=0.3\tau_0$ (before the storage), $t=0.9\tau_0$ (during the storage), and $t=1.5\tau_0$ [after the storage (retrieval)], respectively.
(b)~Surface plots on the evolution of the probe-field intensity in the course of the storage and retrieval of a DSW with large group-velocity dispersion (${\cal D}\approx0.5$), by taking $\rho=|U|^2$ as a function of $x/R_0$, $y/R_0$, and $t/\tau_0$,  at positions $z=0$ (before the storage), $z=2L_{\rm diff}\approx0.94$ mm (during the storage), and $z=4L_{\rm diff}\approx1.88$ mm [after the storage (retrieval)], respectively. }}
\label{fig12}
\end{figure}

Fig.~\ref{fig12} shows the result of numerical simulation on the storage and retrieval of a DSW in a 3D Rydberg gas. For simplicity, we have carried out simulation only on isotropic DSWs (anisotropic DSWs give similar results).
Illustrated in panel (a) are level plots for the evolution of the probe-field intensity $\rho=|q|^2$ in the course of the storage and retrieval of a DSW with ${\cal D}\approx0$, by taking $|q|^2$ as a function of $x$ and $y$, at times $t=0.3\tau_0\approx0.48$ $\mu$s (before the storage), $t=0.9\tau_0\approx1.44$ $\mu$s (during the storage), and $t=1.5\tau_0\approx2.4$ $\mu$s [after the storage (retrieval)], respectively. We see that at $t=0.9\tau_0$ the DSW disappears, which is due to the switched-off of the control field and hence the DSW is stored in the atomic medium. Then at $t=1.5\tau_0$ the DSW reappears, which is due to switched-on of the control field and thus the DSW is retrieved from the atomic medium.

Shown in panel (b) of Fig.~\ref{fig12} are surface plots for the evolution of the probe-field intensity $\rho=|U|^2$ in the course of the storage and retrieval of a DSW with ${\cal D}\approx 0.5$, by taking $|U|^2$ as a function of $x/R_0$, $y/R_0$, and $t/\tau_0$, at positions $z=0$ (before the storage), $z=2L_{\rm diff}\approx0.94$ mm (during the storage), and $z=4L_{\rm diff}\approx1.88$ mm [after the storage (retrieval)], respectively.
One sees that, because of the switched-off and switched-on of the control field, the DSW disappears at $z=2L_{\rm diff}$ and reappears at $z=4L_{\rm diff}$, corresponding to the DSW storage and retrieval in the atomic medium.


The physical reason for the realization of the DSW memory can be understood as an information conversion between the probe pulse and the atomic medium. During the storage, the information carried by the DSW is converted into the atomic spin wave. Then, during the retrieval the information carried by the atomic spin wave is converted back into the DSW. Mathematically, the success of such DSW memory is due to the existence of dark-state polariton allowed by the MB equations (\ref{Max}) and (\ref{Bloch0})~\cite{Fleischhauer2005,Fleischhauer2000}. If the system starts from the dark state $|D\rangle=\Omega_c^\ast|1\rangle-\Omega_p|3\rangle
=\Omega_c^\ast[|1\rangle-(\Omega_p/\Omega_c^\ast)|3\rangle]$, it approximately remains in this dark state, even when $\Omega_c^\ast$ and $\Omega_p$ are approaching zero but their ratio $\Omega_p/\Omega_c^\ast$ keeps nearly to be a constant in the course of the storage and retrieval process~\cite{Chen2014}.

\subsection{Efficiency and fidelity of the DSW memory}

The quality of the storage and retrieval of DSWs can be characterized by efficiency $\eta$ and fidelity $\eta {\mathcal J}$, where $\eta $ and ${\mathcal J}$ are respectively defined by
\begin{subequations}
\begin{align}
\eta & =\frac{\int_{T_{\mathrm{on}}}^{\infty }dt\iint dxdy|\Omega _{p}^{%
\mathrm{retr}}(x,y,t)|^{2}}{\int_{-\infty }^{T_{\mathrm{off}}}dt\iint
dxdy|\Omega _{p}^{\mathrm{stor}}(x,y,t)|^{2}}, \\
{\mathcal J}& =\frac{|\int_{-\infty }^{\infty }dt\iint dxdy\Omega _{p}^{%
\mathrm{retr}}(x,y,t-\Delta T)\Omega _{p}^{\mathrm{stor}}(x,y,t)|^{2}}{%
\int_{-\infty }^{T_{\mathrm{off}}}dt\iint dxdy|\Omega _{p}^{\mathrm{stor}%
}|^{2}\int_{T_{\mathrm{on}}}^{\infty }dt\iint dxdy|\Omega _{p}^{\mathrm{retr}%
}|^{2}},
\end{align}%
\end{subequations}
Here, $\Omega _{p}^{\mathrm{stor}}$ and $\Omega _{p}^{\mathrm{retr}}$ are the stored and retrieved half Rabi frequencies of the probe field, respectively. Based on the results given in Figs.~\ref{fig10} and \ref{fig11}, we estimate the maximum efficiency and fidelity of the DSW memory, which can reach $(\eta, \eta {\mathcal J})\approx (93\%, 90\%)$\, [$(\eta, \eta {\mathcal J})\approx (90\%, 87\%)$ in the 2D (3D) Rydberg gas.
We would like to point that in {\color{cyan}realistic} experiments, because of the decoherence induced by the inhomogeneous broadening due to residual magnetic fields, spin-wave dephasing, and atomic motions~\cite{Hsiao2018}, the maximum efficiency and fidelity of the DSW memory might be lower than the values predicted above.

\begin{figure}
\centering
\includegraphics[width=0.74\columnwidth]{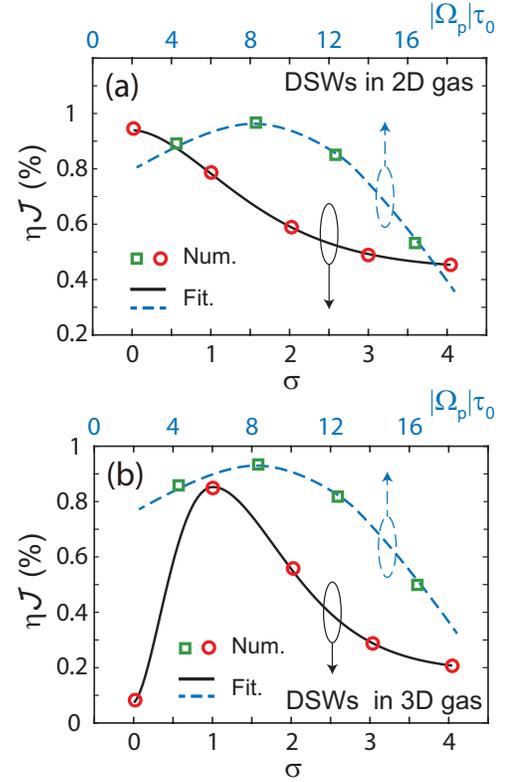}
\caption{{\footnotesize Effects of the strength and nonlocality degree of the Kerr nonlinearity on the quality of the DSW memory in 2D and 3D Rydberg gases.
(a)~The fidelity $\eta{\mathcal J}$ of DSW memory as a function of the probe-field amplitude $|\Omega_p\tau_0|$ (dashed blue line) and the nonlocality degree $\sigma$ (solid black line) in a 2D Rydberg gas.
The symbols (squares and circles) are numerical results, and the dashed and solid curves are the fitting ones.
(b)~The same as in (a), but for the DSWs in a 3D Rydberg gas.
}}
\label{fig13}
\end{figure}

We stress that quality of the DSW memory depends on various physical factors of the system. The first one is the strength of the Kerr nonlinearity.  Fig.~\ref{fig13}(a)
%
%
shows the fidelity $\eta{\mathcal J}$ as a function of the probe-field amplitude $|\Omega_p\tau_0|$ (given by the dashed blue line)
for a 2D Rydberg gas. The symbols (squares) are numerical results, and the dashed curve is the fitting one. It is seen that for moderate amplitudes, $|\Omega _{p}\tau_{0}|\approx 8$, the fidelity reaches its maximum, with the retrieved DSW having nearly the same wave shape
as the original one prior to the storage. For small and large amplitudes, the fidelity features small values, implying that the retrieved DSW suffers evident distortion. This happens because, for the weak and strong probe-field amplitudes, the Kerr nonlinearity is either too weak or too strong to interplay with the diffraction of the DSW. The fidelity $\eta{\mathcal J}$ as a function of $|\Omega _{p}\tau_{0}|$ for the DSW in a 3D gas is also calculated, given by the dashed blue line in Fig.~\ref{fig13}(b), which has the similar features depicted above.

The nonlocality degree of the Kerr nonlinearity can also lead to significant effect on the quality of the DSW memory. The solid black line in Fig.~\ref{fig13}(a) is the fidelity $\eta{\mathcal J}$ as a function of the nonlocality degree $\sigma$ for the DSW in the 2D gas (the circles are numerical results and the solid curve is the fitting one). One sees that $\eta{\mathcal J}$ decreases when $\sigma$ increases. The dependence of $\eta{\mathcal J}$ on $\sigma$ for the DSW in a 3D gas is also obtained [given by the solid black line
in Fig.~\ref{fig13}(b)], which is sharply different from the case of the DSW in the 2D gas. In particular, $\eta{\mathcal J}$ for the DSW memory in the 3D gas reaches its maximum at a moderate value of the nonlocality degree, i.e. $\sigma\approx 1$. However, when $\sigma$ is deviated from this moderate value, $\eta{\mathcal J}$  is reduced rapidly. The physical reason for such phenomenon is the following. For small $\sigma$, the system works in the weak nonlocality regime where the DSW in the 3D gas is unstable; on the other hand, for large $\sigma$,
the Rydberg blockade effect can take effect, leading to the deterioration of EIT and hence a large absorption of the DSW.

\section{Summary}\label{section7}

In this work, we have proposed and analyzed in detail a scheme for generating weak-light DSWs and realizing their active manipulations by means of the giant nonlocal Kerr nonlinearity in cold Rydberg atomic gases working under the condition of EIT. We have shown that in 2D Rydberg gases even a very weak nonlocality of the Kerr nonlinearity can significantly change the edge speeds of DSWs, which display a singular behavior when the local sound speed ($c_s$) approaches a critical value ($c_{cr}$). Moreover, the weak nonlocality may induce the instability of DSWs when $c_s$ reaches or overpasses $c_{cr}$.  However, as the local sound speed increases, the increase of the edge speeds of DSWs becomes much slower in the intermediate nonlocality regime, where the singular behavior of edge speeds disappears and the instability of DSWs is thoroughly suppressed.

We have demonstrated that in a 3D Rydberg gas DSWs can also be generated, which are stable during propagation when the system works in the intermediate nonlocality regime. Based on the EIT effect and the giant nonlocal Kerr nonlinearity due to the Rydberg-Rydberg interaction, DSWs found here have extremely low generation power ($\le 10$ nano-watts). We have also demonstrated that the active control of such DSWs can be effectively implemented; especially, they can be stored and retrieved with high efficiency and fidelity through switching-off and switching-on of the control laser field. Our analytical and numerical study paves a route to create, manipulate, store, and retrieve DSWs in strongly interacting Rydberg gases. Such active controllability in this setting may be useful for exploring intriguing physics of DSWs~\cite{Simmons2020,Marcucci2019APX} and developing optical technologies based on nonlinear and nonlocal Rydberg media~\cite{Adams2020}.

\vspace{5mm}
\acknowledgments

C. H., Z. B., and G. H. acknowledge the National Natural Science Foundation of China (NSFC) under Grant Nos.~11974117, 11904104, and 11975098, the National Key Research and Development Program of China under Grant Nos.~2017YFA0304201, and the Shanghai Pujiang Program under grant No. 21PJ1402500; W. L. acknowledges support from the EPSRC through Grant No. EP/W015641/1.

\appendix

\section{Optical Bloch equations}\label{appendixA}


The optical Bloch equation describing the time evolution of the density-matrix elements $\rho_{\alpha\beta}\equiv\langle{\hat S}_{\beta\alpha}\rangle$ reads
\begin{subequations}\label{Bloch}
\begin{align}
&i\frac{\partial}{\partial t}\rho_{11}-i\Gamma_{12}\rho_{22}+\Omega_{p}^*\rho_{21}-\Omega_{p}\rho_{12}=0, \\
&i\frac{\partial}{\partial t}\rho_{22}+i\Gamma_{12}\rho_{22}-i\Gamma_{23} \rho_{33}+\Omega_c^*\rho_{32}-\Omega_{c}\rho_{23}\notag\\
&\qquad-\Omega_{p}^*\rho_{21}+\Omega_{p}\rho_{12}=0 , \\
&i\frac{\partial}{\partial t}\rho_{33}+i\Gamma_{23} \rho_{33}-\Omega_c^*\rho_{32}+\Omega_c\rho_{23}=0, \\
&\left(i\frac{\partial }{\partial t}+d_{21}\right)\rho _{21}+\Omega_c^{*}\rho_{31}-\Omega_{p}(\rho_{22}-\rho_{11})=0,\\
&\left(i\frac{\partial }{\partial t}+d_{31}\right)\rho_{31}-\Omega_{p}\rho_{32}+\Omega_c\rho_{21}\notag\\
&\qquad-\mathcal N_a\int d^{3}{\bf r^{\prime}}  V({\bf r^{\prime}}-{\bf r}) \rho_{33,31}({\bf r^{\prime}},{\bf r},t)=0,\label{rho31}\\
&\left(i\frac{\partial }{\partial t}+d_{32}\right)\rho _{32}-\Omega_{p}^*\rho_{31}-\Omega_{c}(\rho_{33}-\rho_{22})\notag\\
&\qquad-\mathcal N_a\int d^{3}{\bf r^{\prime}}  V({\bf r^{\prime}}-{\bf r}) \rho_{33,32}({\bf r^{\prime}},{\bf r},t)=0,\label{rho32}
\end{align}
\end{subequations}
where $\rho_{\alpha\beta}=\langle \hat{S}_{\alpha\beta}\rangle$ is one-body density matrix element, $d_{21}=\Delta_2+i\gamma_{21}$, $d_{31}=\Delta_3+i\gamma_{31}$, $d_{31}=\Delta_3+i\gamma_{31}$,
$d_{32}=\Delta_3-\Delta_2+i\gamma_{32}$, $\gamma_{\alpha\beta}=(\Gamma_{\alpha} + \Gamma_{\beta})/2 + \gamma_{\alpha\beta}^{\rm dep}$ ($\alpha\neq \beta$; $\Gamma_{1}=0$), and $\Gamma_{\beta}=\sum_{\alpha<\beta}
\Gamma_{\alpha\beta}$, with $\Gamma_{\alpha\beta}$ the spontaneous emission decay rate and $\gamma_{\alpha\beta}^{\rm dep}$ the dephasing rate from $|\beta\rangle$ to $|\alpha\rangle$. For cold atoms,
$\gamma_{\alpha\beta}^{\rm dep}$ is usually much less than $\Gamma_{\alpha\beta}$ and hence is negligible.

The last terms on the left hand side of Eqs.~(\ref{rho31}) and~(\ref{rho32})] include the two-body correlators $\rho_{33,3\alpha}({\bf r',r},t)\equiv \langle{\hat S}_{33}({\bf r'},t){\hat S}_{3\alpha}({\bf
r},t)\rangle$ $(\alpha=1,2)$, which are contributed by the Rydberg-Rydberg interaction.

\section{Dark-soliton solutions}\label{Appendix2}

If the diffraction is significant enough to balance the Kerr nonlinearity, it is possible to have dark solitons in the present system since the Kerr nonlinearity is strong and defocusing. The dark soliton solutions of Eq.~(\ref{Euler1}) in the weak nonlocality regime can be found by using the traveling-wave method, i.e. via the combination $\xi-V_s\zeta$ with $V_s$ the velocity of solitons.

We assume that far from the soliton location the flow velocity of the light fluid vanishes and the light intensity approaches to
the value of background $\rho_b$. Eqs.~(\ref{Euler1a}) and (\ref{Euler2}) support the following dark soliton solution
\begin{eqnarray}\label{soliton}
\xi=&&2\sqrt{\frac{\cR}{\cG}}\arccot\left(\frac12\sqrt{\frac{\cG(1-4\cR\rho)}{\cR(\cG\rho-V_s^2)}}\right)-\frac12\sqrt{\frac{1-4\cR\rho_b}{\cG\rho_b-V_s^2}} \notag\\
&&\times\ln\frac{\sqrt{\Theta/\Xi}-1}{\sqrt{\Theta/\Xi}+1},
\end{eqnarray}
where $\Theta=(\cG\rho_b-V_s^2)(1-4\cR\rho)$ and $\Xi=(\cG\rho-V_s^2)(1-4\cR\rho_b)$. 
The intensity at the center of the soliton is $\rho(\xi=0)=\rho_m=V_s^2/\cG$, and the depth of the dark soliton reads
\begin{equation}\label{soliton_depth1}
\rho_b-\rho_m=\rho_b-V_s^2/\cG,
\end{equation}
which depends on the soliton velocity $V_s$. When the soliton is stationary ($V_s=0$), it has zero intensity at the soliton center, corresponding to the largest depth. Far from the soliton location, $|\xi|\to\infty$, the soliton
follows the asymptotic behavior
$\rho_b-\rho\propto e^{-\tk|\xi|}$, with
$\tk=2\sqrt{(\cG\rho_b-V_s^2)/(1-4\cR\rho_b)}$,
which determines the inverse half-width of the soliton. As a result,
the soliton velocity can be expressed as
\begin{equation}\label{soliton_velocity1}
V_s=\sqrt{\cG\rho_b-\left(\frac{1}{4}-\cR\rho_b\right)\tk^2},
\end{equation}
which is related to the dispersion relation of linear wave (\ref{dispersion}) by
\begin{equation}\label{soliton_velocity2}
V_s=\frac{\om(ik)}{ik}=\frac{\tom(\tk)}{\tk},
\end{equation}
with
\begin{equation}\label{tom0}
\tilde{\omega}=\tk\sqrt{\cG\rho_b-\left(\frac{1}{4}-\cR\rho_b\right)\tk^2}.
\end{equation}
This is a consequence of Stokes' remark~\cite{Stokes1905} that the soliton tails are described by the corresponding linear equation and propagate with the same velocity as the soliton itself; hence Eq.~(\ref{tom0}) can also
be obtained from the dispersion relation (\ref{dispersion}) by means of the replacements $k\to ik$ and $\om\to i\om$.

Shown in Figure~\ref{fig2}(b) of the main text is the soliton intensity profile $\rho$ with different values of $({\cR}, V_s)$ as a function of $\xi$.
We stress that, though a model similar to (\ref{NLS2}) and related solutions have been considered in Ref.~\cite{kb-2000,tsoy-10}, the nonlocal dark solitons presented here have much lower generation power and they are more flexible for controls because in the present the system the Kerr nonlinearity is extremely large and its nonlocality degree can be manipulated actively.


\begin{thebibliography}{99}
\bibitem{note-blockade} Because of the strong dipole-dipole interaction between Rydberg excitations, only one excitation is allowed within a volume known
as the blockade sphere. The radius of the blockade sphere is referred to as the Rydberg blockade radius.
\bibitem{note-average}The definition of the average of the operator $\hat{O}$ is given by $\langle G|\hat{O}|G\rangle$, with $|G\rangle$ is the ground state of the atomic ensemble when the probe field is not applied, i.e. $|G\rangle=|1,1,...,1\rangle$.
\bibitem{note-2D-NNLSE}
When both the first- and second-order dispersions can be neglected, i.e. $1/\lambda\ll 1 $ and ${\cal D}\ll 1$, Eq.~(\ref{NLS1}) can also reduced into
the 2D NNLSE (\ref{NLS12}). In this case, the probe field is a stationary light beam. Such approach is valid for the probe-field time duration, $\tau_0$, is long enough (e.g. $\tau_0\ge\, 5\,\mu$s).
\bibitem{notediffraction}
Since the second term on the right hand side of Eq.~(\ref{NLS2}) represents a diffractive effect, the SWs obtained here are actually diffractive ones. But, as widely used in literature, we still call them DSWs for simplicity.
\end{thebibliography}
\end{document}